\documentclass[aps,twocolumn,showpacs,amsmath,amsfonts]{revtex4-1}

\usepackage{graphicx}
\usepackage{bm,units}

\newcommand{\deriv}[2]{\frac{\partial #1}{\partial #2}}
\newcommand{\ii}{\mathrm{i}}
\newcommand{\ee}{\mathrm{e}}
\newcommand{\dd}{\mathrm{d}}
\newcommand{\scatt}{a_{\mathrm{sc}}}
\newcommand{\scc}{\mathrm{sc}}

\begin{document}

\title{Variational methods with coupled Gaussian functions for Bose-Einstein 
condensates with long-range interactions. II. Applications}

\author{Stefan Rau}
\author{J\"org Main}
\author{Holger Cartarius}
\author{Patrick K\"oberle}
\author{G\"unter Wunner}
\affiliation{Institut f\"ur Theoretische Physik 1, Universit\"at Stuttgart,
  70550 Stuttgart, Germany}
\date{\today}

\begin{abstract}
Bose-Einstein condensates with an attractive $1/r$ interaction and with 
dipole-dipole interaction are investigated in the framework of the Gaussian
variational ansatz introduced by S.~Rau, J.~Main, and G.~Wunner 
[Phys.\ Rev.\ A, submitted].
We demonstrate that the method of coupled Gaussian wave packets is a 
full-fledged alternative to direct numerical solutions of the 
Gross-Pitaevskii equation, or even superior in that coupled Gaussians are 
capable of producing both, stable and unstable states of the Gross-Pitaevskii 
equation, and thus of giving access to yet unexplored regions of the space 
of solutions of the Gross-Pitaevskii equation.  As an alternative to 
numerical solutions of the Bogoliubov-de Gennes equations, the stability of 
the stationary condensate wave functions is investigated by analyzing the 
stability properties of the dynamical equations of motion for the Gaussian
variational parameters in the local vicinity of the stationary fixed points.
For blood-cell-shaped dipolar condensates it is shown that on the route to 
collapse the condensate passes through a pitchfork bifurcation, where 
the ground state itself turns unstable, before it finally vanishes in a 
tangent bifurcation.
\end{abstract}


\maketitle

\section{Introduction}
\label{sec:intro}
In the previous paper \cite{paper1} variational methods with coupled Gaussian 
functions for Bose-Einstein condensates with long-range interactions have
been developed.
The purpose of this paper is to demonstrate the power of this approach
by applying the variational methods to two different types of condensates,
viz.\ a monopolar condensate with an attractive gravity-like $1/r$ interaction
and a dipolar condensate.

Monopolar condensates with an attractive $1/r$ interaction, originally 
proposed by O'Dell et al.\ \cite{ODellPRL84}, have unique stability properties.
For a wide range of the scattering length the condensate is stable without an
external trap.
Additionally, the gravity-like interaction of a monopolar condensate may be 
an opportunity to investigate usually large-scale physical properties like,
e.g., boson stars \cite{RuffiniPhysRev187} at a laboratory scale. 
The ``monopolar'' interaction of two neutral atoms with positions $\bm r$ and 
$\bm r'$ is induced by the presence of external electromagnetic radiation.
O'Dell et al.\ \cite{ODellPRL84} suggest six triads of orthogonal laser beams
to induce the interatomic potential
$W_{\mathrm{lr}}(\bm r,\bm r')=-u/|\bm r-\bm r'|$,
where $u$ depends on the intensity and wave vector of the radiation, and on
the isotropic polarizability of the atoms.
Although this system has not yet been realized experimentally it has already
served as a model to compare results obtained analytically and with exact
numerical techniques \cite{PapadopoulosPRA76,holgerPRA77,holgerPRA78}.

By contrast, Bose-Einstein condensates with a long-range dipole-dipole
interaction
$W_{\mathrm{lr}}(\bm r,\bm r') \sim (1-3\cos^2\theta)|\bm r-\bm r'|^{-3}$
have been obtained experimentally with $^{52}\mathrm{Cr}$ atoms in 2005 
by Griesmaier et al.\ \cite{EXGriesmaierPRL94,EXStuhlerPRL95}, and more 
recently in 2008 by Beaufils et al.\ \cite{BeaufilsPRA77}.
The collapse of the condensate also has been subject to extensive experimental 
studies \cite{Koch08Nature}.
Theoretical investigations have so far mostly been based on lattice 
simulations \cite{Ronen,Ronen06a,WilsonPRA80} or on a simple variational
approach with a Gaussian type orbital \cite{gelbPatrick}.
For a review on dipolar condensates see \cite{Lahaye2009}.

In this paper we extend and elaborate in more detail preliminary work
presented in \cite{Rau10}.
In Sec.~\ref{chap:monopolar_interaction} the results for the monopolar 
condensates are presented and discussed.
It is shown that only three to five coupled Gaussians are sufficient to 
achieve convergence of the mean-field energy, the chemical potential, 
and the lowest eigenvalues of the stability matrix.
In Sec.~\ref{chap:dipolar_interaction} the method of coupled Gaussians 
is applied to dipolar BEC to clarify the theoretical nature of the collapse 
mechanism of blood-cell shaped condensates.
On the route to collapse the condensates passes through a pitchfork 
bifurcation, where the ground state itself turns unstable, before it 
finally vanishes in a tangent bifurcation.
Conclusions are drawn in Sec.~\ref{sec:conclusion}.

\section{Monopolar Condensates}
\label{chap:monopolar_interaction}
The time-independent Gross-Pitaevskii equation (GPE) 
for a self-trapped condensate with a short-range contact interaction with 
scattering length $a$ and a long-range monopolar interaction reads
\begin{align}
   \bigg[ -\Delta + 8 \pi N\frac{a}{a_u} \left | \psi(\bm{r})\right |^2 
   - 2N \int \mathrm{d}^3 \bm{r}' \frac{\left | \psi(\bm{r}')\right |^2} 
 {\left | \bm{r} - \bm{r}' \right |} &\bigg]  \psi(\bm r)\nonumber \\
 = \mu &\psi(\bm r) \; ,
\label{eq:monopolarGPE}
\end{align}
where the natural ``atomic'' units introduced in Ref.\ \cite{PapadopoulosPRA76} 
were used.
Lengths are measured in units of a ``Bohr radius'' $a_u=\hbar^2/(m u)$,
energies in units of a ``Rydberg energy'' $E_u=\hbar^2/(2 m a_u^2)$,
and times in units of $t_u = \hbar/E_u$, where $u$ determines the strength of
the atom-atom-interaction \cite{ODellPRL84}, and $m$ is the mass of one boson.
The number of bosons $N$ can be eliminated from Eq.~\eqref{eq:monopolarGPE}
by using scaling properties of the system \cite{PapadopoulosPRA76,holgerPRA78}.
We define
$\bm{ r} = \tilde{\bm r}/N$, $\psi = N^{3/2}\tilde \psi$,
which leaves the norm of the wave function invariant,
$\scatt=N^2a/a_u$, $\tilde\mu=\mu/N^2$,
omit the tilde in what follows, substitute $\mu\to \ii(\dd/\dd t)$, and
finally obtain the time-dependent GPE in scaled ``atomic'' units
\begin{align}
   \bigg[ &-\Delta + 8 \pi \scatt \left | \psi(\bm{r},t)\right |^2 \nonumber \\
   &- 2 \int \mathrm{d}^3 \bm{r}' \frac{\left | \psi(\bm{r}',t)\right |^2} 
 {\left | \bm{r} - \bm{r}' \right |} \bigg]  \psi(\bm r,t)
 = \ii \frac{\mathrm{d}}{\mathrm{d}t} \psi(\bm r,t) \; .
\label{eq:GPE_mono}
\end{align}
It is known that for Bose-Einstein condensates with attractive $1/r$
interaction two real radially symmetric solutions, the ground state and
a collectively excited state, exist. By varying the scattering length
$a_\mathrm{sc}$, they are created in a tangent bifurcation at a critical
value of $a_\mathrm{sc}$ \cite{PapadopoulosPRA76,holgerPRA77,holgerPRA78}. 
Below the tangent bifurcation no stationary solutions exist. 
Approaching the bifurcation point from higher scattering lengths the 
condensate collapses. 
A similar behavior is found for dipolar condensates \cite{gelbPatrick}
and condensates without long-range interactions \cite{HuepePRA}.

For the monopolar GPE numerically exact solutions exist.
The numerical procedure for the direct integration of the GPE is introduced
in Refs.\ \cite{PapadopoulosPRA76,holgerPRA77,holgerPRA78}.
It is our purpose to investigate the two states connected via the 
tangent bifurcation with the coupled Gaussian wave packet method.
We use the ansatz
\begin{equation}
\label{eq:coupledAnsatz}
 \psi(r) = \sum_{k=1}^N \ee^{\ii(a^k r^2 + \gamma^k)} = \sum_{k=1}^N g^k
\end{equation}
for condensates with spherical symmetry.
Following the procedure outlined in \cite{paper1} the dynamical equations
for the variational parameters read
\begin{subequations}
\label{eq:eomReducedSymmetry_r}
\begin{align}
 \dot\gamma^k &= 6 \ii a^k - v_0^k \; , \\
 \dot a^k &= - 4 (a^k)^2 - \frac{1}{2} V_2^k \; ,
\end{align}
\end{subequations}
for $k=1,\dots,N$.
The quantities $v_0^k$ and $V_2^k$ are obtained from the 
$(2N\times 2N)$-dimensional linear set of equations ($k,l=1,\dots,N$)
\begin{equation}
\begin{pmatrix}
 (1)_{kl} &    (r^2)_{kl} \\[0.5ex]
 (r^2)_{lk}   &  (r^4)_{kl}
\end{pmatrix} 
\begin{pmatrix}
 v_0^k \\[0.5ex]
 \frac{1}{2} V_2^k
\end{pmatrix}
 = \sum_{k = 1}^{N}	
\begin{pmatrix}
 \langle g^l|V_{\rm{eff}}|g^k\rangle\\[1ex]
 \langle g^l|r^2 V_{\rm{eff}}|g^k\rangle
\end{pmatrix},
\end{equation}
where $V_\mathrm{eff} = V_\scc + V_\mathrm{m}$ is the sum of the contact and 
monopolar potential, and the matrix elements read
\begin{subequations}
\begin{align}
& \left( 1 \right)_{lk}: \bigl< g^l \big| g^k \bigr> 
 	=\pi^{\nicefrac{3}{2}} \dfrac{\ee^{\ii \gamma^{kl}}}{\left(- \ii  a^{kl}  \right)^{\nicefrac{3}{2}}}\; ,\\
& \left( r^2 \right)_{lk} :  \bigl< g^l \bigl| r^2 \bigr| g^k \bigr> 
 	=  \dfrac{3}{2} \pi^{\nicefrac{3}{2}} \dfrac{ \ee^{\ii \gamma^{kl}}}{ \left( -\ii a^{kl} \right)^{\nicefrac{5}{2}}}\; ,\\
& \left( r^4 \right)_{lk} : \bigl< g^l \bigl| r^4 \bigr| g^k \bigr> 
 	=   \dfrac{15}{4} \pi^{\nicefrac{3}{2}} \dfrac{ \ee^{\ii \gamma^{kl}}}{ \left( -\ii a^{kl} \right)^{\nicefrac{7}{2}}}\; ,\\
& \bigl< g^l \bigl| V_\scc \bigr| g^k \bigr> 
=\sum_{i,j = 1}^{N} 8 \scatt \pi^{\nicefrac{5}{2}} 
\dfrac{\ee^{\ii \gamma^{klij}}}{\left(- \ii  a^{klij}  \right)^{\nicefrac{3}{2}}}\; ,\\
&\bigl< g^l \bigl| V_\mathrm{m} \bigr| g^k \bigr> 
= 4\sum_{i,j = 1}^{N}  \pi^{\nicefrac{5}{2}} \dfrac{\ee^{\ii{\gamma^{klij}}} }{a^{ij}a^{kl}\sqrt{- \ii a^{klij}}}\; ,\\
& \bigl< g^l \bigl| r^2 V_\scc \bigr| g^k \bigr> 
= 12 \sum_{i,j = 1}^{N}  \scatt \pi^{\nicefrac{5}{2}} 
\dfrac{\ee^{\ii \gamma^{klij}}}{\left(- \ii  a^{klij}  \right)^{\nicefrac{5}{2}}}\; ,\\
& \bigl< g^l \bigl| r^2 V_\mathrm{m} \bigr| g^k \bigr> 
=2 \sum_{i,j = 1}^{N} \pi^{\nicefrac{5}{2}} \dfrac{\ee^{\ii{\gamma^{klij}}}  \left(2 a^{ij} + 3 a^{kl} \right)}
{a^{ij}\left(a^{kl}\right)^2     \left(- \ii a^{klij} \right)^{\nicefrac{3}{2}}}\; ,
\end{align}
\end{subequations}
with the abbreviations
\begin{subequations}
\label{def:abbreviations_monopolar}
\begin{align}
\label{def:abbreviations}
 &a^{kl} = a^k - \left(a^l\right)^*\; , \quad
 a^{ij} = a^i - \left(a^j\right)^*\; ,\\
 &\gamma^{kl} = \gamma^k - \left(\gamma^l\right)^*\; , \quad
 \gamma^{ij} = \gamma^i - \left(\gamma^j\right)^*\; ,\\
 &a^{klij} = a^{kl} + a^{ij}\; , \quad
 \gamma^{klij} = \gamma^{kl} + \gamma^{ij}\; .
\end{align}
\end{subequations}

\subsection{Stationary solutions}
\label{subsection:stationarySolutions_coupled_monopolar}
Stationary solutions of the GPE can be computed via a nonlinear root search 
of the dynamical equations \eqref{eq:eomReducedSymmetry_r} as outlined
in \cite{paper1}.
The Gaussian parameters obtained from the root search are then used to 
calculate the mean field energy
\begin{align}
 E_{\rm mf} &= \sum_{k,l=1}^N 2 \pi^{\nicefrac{3}{2}}\ee^{\ii \gamma^{kl}}
\Biggl[
-3 \frac{a^k \left(a^l\right)^*}{\left(a^{kl}\right)^2\sqrt{-\ii a^{kl}}} \nonumber \\*
&+ \sum_{i,j=1}^N \frac{\pi \ee^{\ii \gamma^{ij}}} {\sqrt{-\ii a^{klij} }}
\left( \frac{2 \scatt}{-\ii a^{klij}} + \frac{1}{a^{ij}a^{kl}}
\right)
\Biggr]
\end{align}
and the chemical potential
\begin{align}
 \mu &= \sum_{k,l=1}^N 2 \pi^{\nicefrac{3}{2}}\ee^{\ii \gamma^{kl}}
\Biggl[
-3 \frac{a^k \left(a^l\right)^*}{\left(a^{kl}\right)^2\sqrt{-\ii a^{kl}}} \nonumber \\*
&+ \sum_{i,j=1}^N \frac{\pi \ee^{\ii \gamma^{ij}}} {\sqrt{-\ii a^{klij} }}
\left( \frac{4 \scatt}{-\ii a^{klij}} + \frac{2}{a^{ij}a^{kl}}
\right)
\Biggr] \; .
\end{align}
For the variational ansatz with only one {\em single} Gaussian function ($N=1$)
the results can be obtained analytically \cite{holgerPRA77,holgerPRA78}, viz.
\begin{equation}
\label{eq:E1g}
   E_{\mathrm{mf}}^{N=1}
 = -\frac{4}{9 \pi}\frac{1 \pm 2 \sqrt{1+\frac{8 \scatt}{3 \pi}}}
   {\left(1 \pm \sqrt{1+\frac{8 \scatt}{3 \pi}}\right)^2}
\end{equation}
for the mean field energy, and
\begin{equation}
\label{eq:mu1g}
 \mu^{N=1} = -\frac{4}{9 \pi}\frac{5 \pm 4 \sqrt{1+\frac{8 \scatt}{3 \pi}}}
  {\left(1 \pm \sqrt{1+\frac{8 \scatt}{3 \pi}}\right)^2}
\end{equation}
for the chemical potential.

\begin{figure}
\includegraphics[width=0.85\columnwidth]{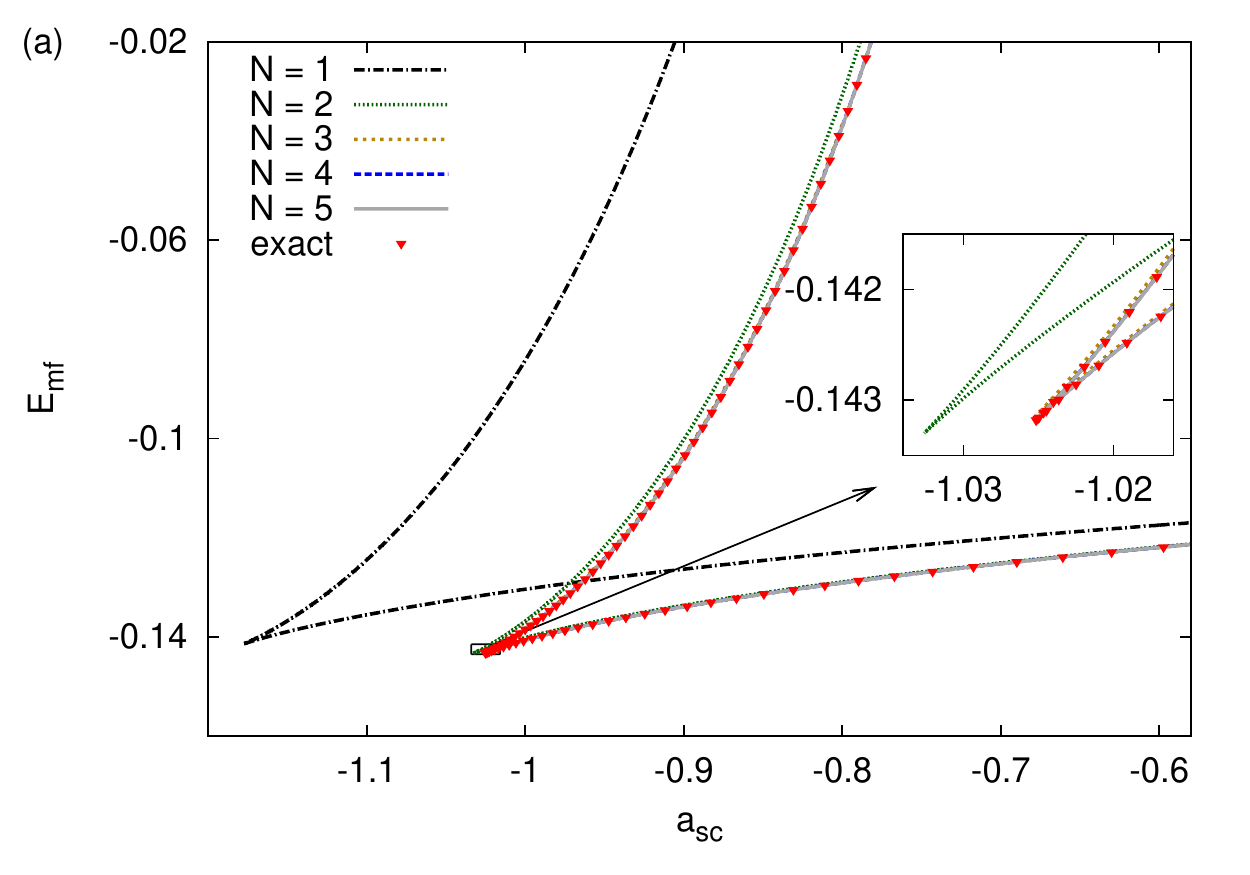}
\includegraphics[width=0.85\columnwidth]{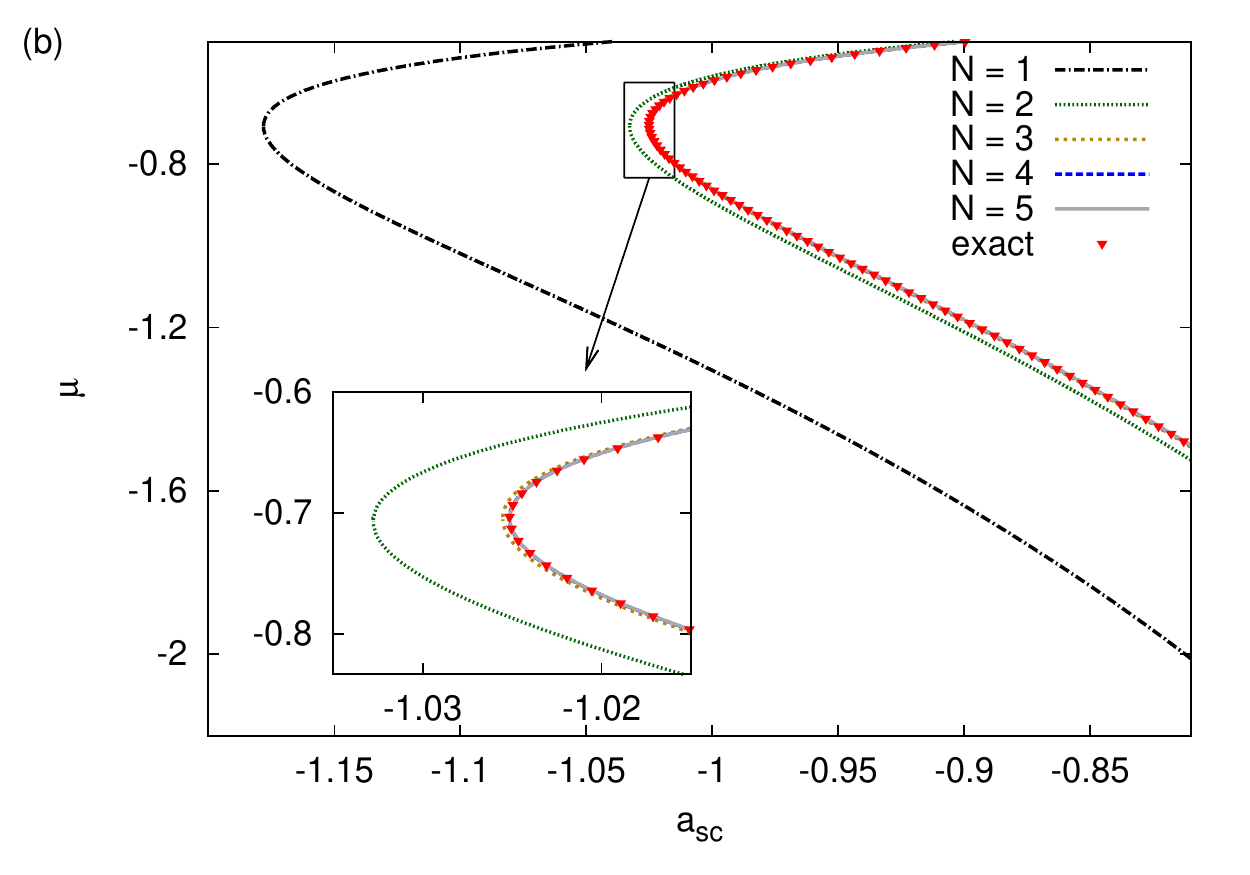}
\caption{\label{fig:monopolar}
 (a) Mean field energy $E_{\mathrm{mf}}$ for
 self-trapped condensates with attractive $1/r$ interaction as a function
 of the scattering length obtained by using up to five coupled Gaussian
 wave packets ($N = 1,\dots, 5$, respectively) in comparison with the result
 of the accurate numerical solution of the stationary Gross-Pitaevskii
 equation (exact).
 (b) Same as (a) but for the chemical potential $\mu$.}
\end{figure}

The results of the variational computations with $N=1$ to $N=5$ coupled
Gaussian functions are presented in Fig.~\ref{fig:monopolar}(a) for the
mean field energy and in (b) for the chemical potential.
For comparison the results of the exact solution obtained by direct 
numerical integration of the GPE are shown as red triangles.
Although the variational solution with a single Gaussian significantly
differs from the exact calculation, the qualitative behavior is the same: 
two solutions emerge in a tangent bifurcation at a critical scattering 
length, which is $\scatt^\mathrm{cr,v}=-3\pi/8=-1.178097$ and 
$\scatt^\mathrm{cr,n}=-1.025147$ for the variational and exact calculation,
respectively \cite{PapadopoulosPRA76}.
Note that the excited state with higher mean field energy has a lower 
chemical potential than the ground state.

The main purpose of Fig.~\ref{fig:monopolar} is to demonstrate how the 
coupling of only $N=2$ to $N=5$ Gaussian functions drastically improves 
the results obtained with a simple Gaussian type orbital.
The coupling of only two Gaussian functions already leads to a significant 
improvement for both, the mean field energy value and the chemical potential.
For three or more coupled Gaussians, the results in 
Fig.~\ref{fig:monopolar} can no longer be distinguished from the numerically 
exact solution.
The enlargements in Fig.~\ref{fig:monopolar} illustrate the rapid 
convergence of the results with increasing number $N$ of coupled Gaussians.

Detailed values for the mean field energy and the chemical potential of the
ground state and the excited state at scattering length $\scatt=-1$ close to
the tangent bifurcation are presented in Table~\ref{tab1}.
\begin{table}[b]
\caption{\label{tab1}
 Dependence of the critical scattering length $\scatt^\mathrm{cr}$ of the
 tangent bifurcation and the mean field energy and the chemical potential
 of the ground state (g) and the excited state (e) at scattering length
 $\scatt=-1$ on the number $N$ of coupled Gaussian functions.}
\begin{tabular}{c|ccccc}
$N$ &$\scatt^\mathrm{cr}$ & $E_{\rm mf}^{\rm g}$ & $\mu^{\rm g}$  & $E_{\rm mf}^{\rm e}$  & $\mu^{\rm e}$\\
\hline
$1$     & $-1.178097$ & $-0.130383$ & $-0.480807$ & $-0.084219$ & $-1.304592$\\
$2$     & $-1.032780$ & $-0.140151$ & $-0.584799$ & $-0.136826$ & $-0.892376$\\
$3$     & $-1.025527$ & $-0.140637$ & $-0.595154$ & $-0.138380$ & $-0.862562$\\
$4$     & $-1.025167$ & $-0.140655$ & $-0.595659$ & $-0.138449$ & $-0.860862$\\
$5$     & $-1.025149$ & $-0.140656$ & $-0.595682$ & $-0.138452$ & $-0.860767$\\
$\infty$& $-1.025147$ & $-0.140657$ & $-0.595685$ & $-0.138453$ & $-0.860757$
\end{tabular}
\end{table}
The coupling of three Gaussian functions already yields an accuracy of more 
than four digits for the mean field energy of the ground state.
Using up to five Gaussians, we can safely assume the variational result
as to be fully converged to the exact numerical computation marked as 
$N=\infty$.
The rapid convergence of the critical scattering length $\scatt^\mathrm{cr}$
with increasing number of coupled Gaussians is also illustrated 
in Table~\ref{tab1}.

In previous work \cite{PapadopoulosPRA76} it has been noted that 
the peak amplitude of the exact wave function at the center $r=0$ is 
poorly reproduced by a Gaussian type orbital.
Here, we illustrate that superpositions of Gaussian functions can accurately 
approximate the exact wave function and thus also provide the correct peak 
amplitude.

\begin{figure}
\includegraphics[width=0.85\columnwidth]{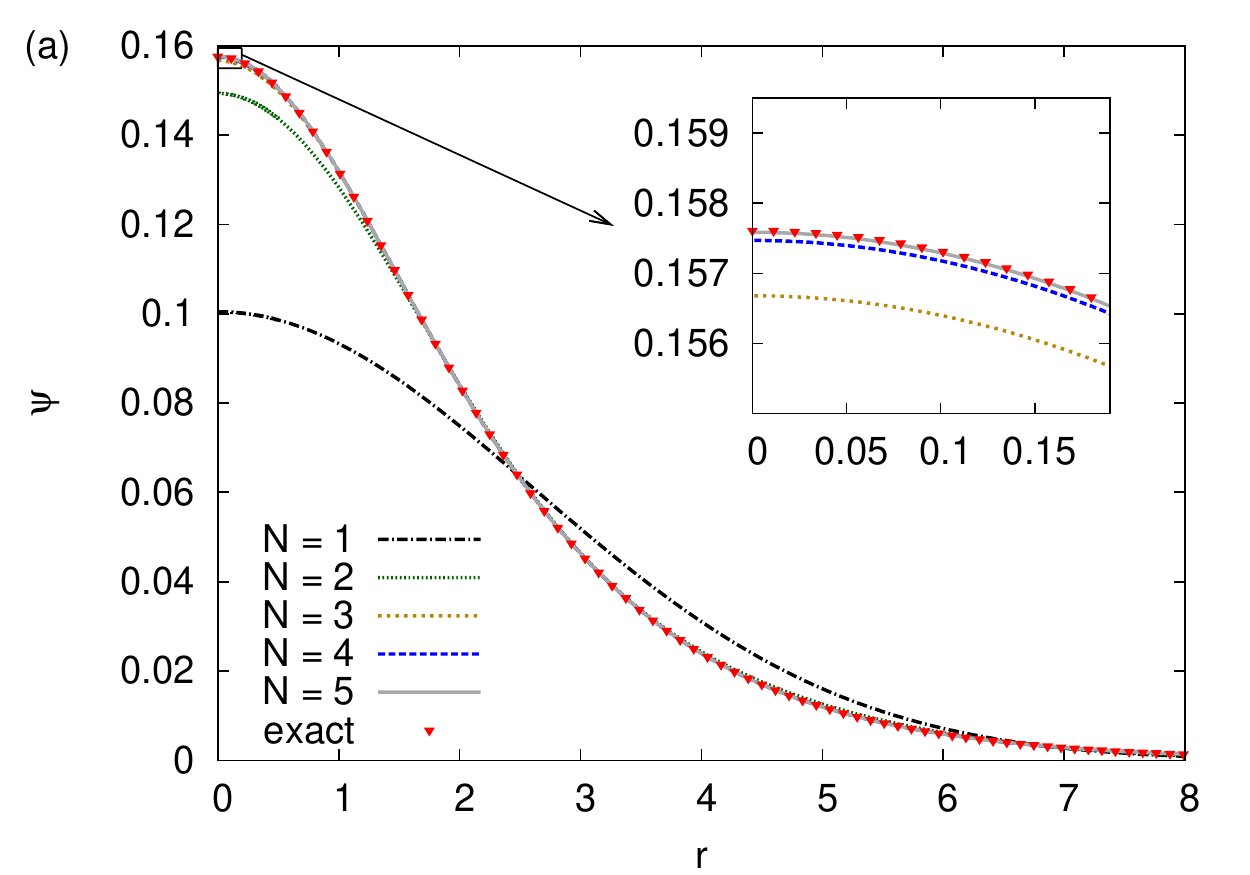}
\includegraphics[width=0.85\columnwidth]{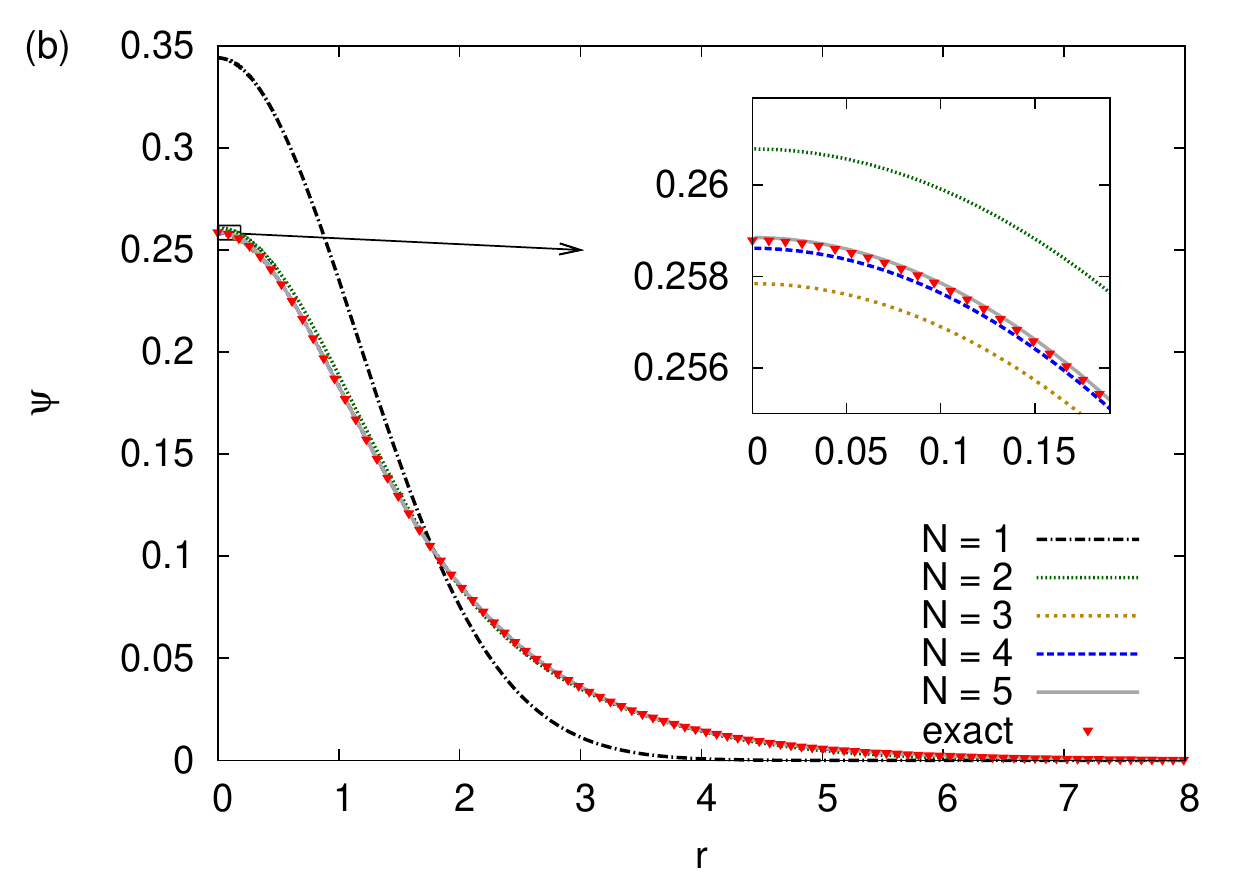}
\caption{\label{fig:Psi_12345_b=-1}
 Wave function $\psi(r)$ of (a) the stable ground state, (b) the excited state,
 at scattering length $\scatt = -1$ close to the exact numerical bifurcation.
 The calculations with coupled Gaussians converge rapidly to the exact 
 numerical wave function (triangles).}
\end{figure}
\begin{figure}
\includegraphics[width=0.85\columnwidth]{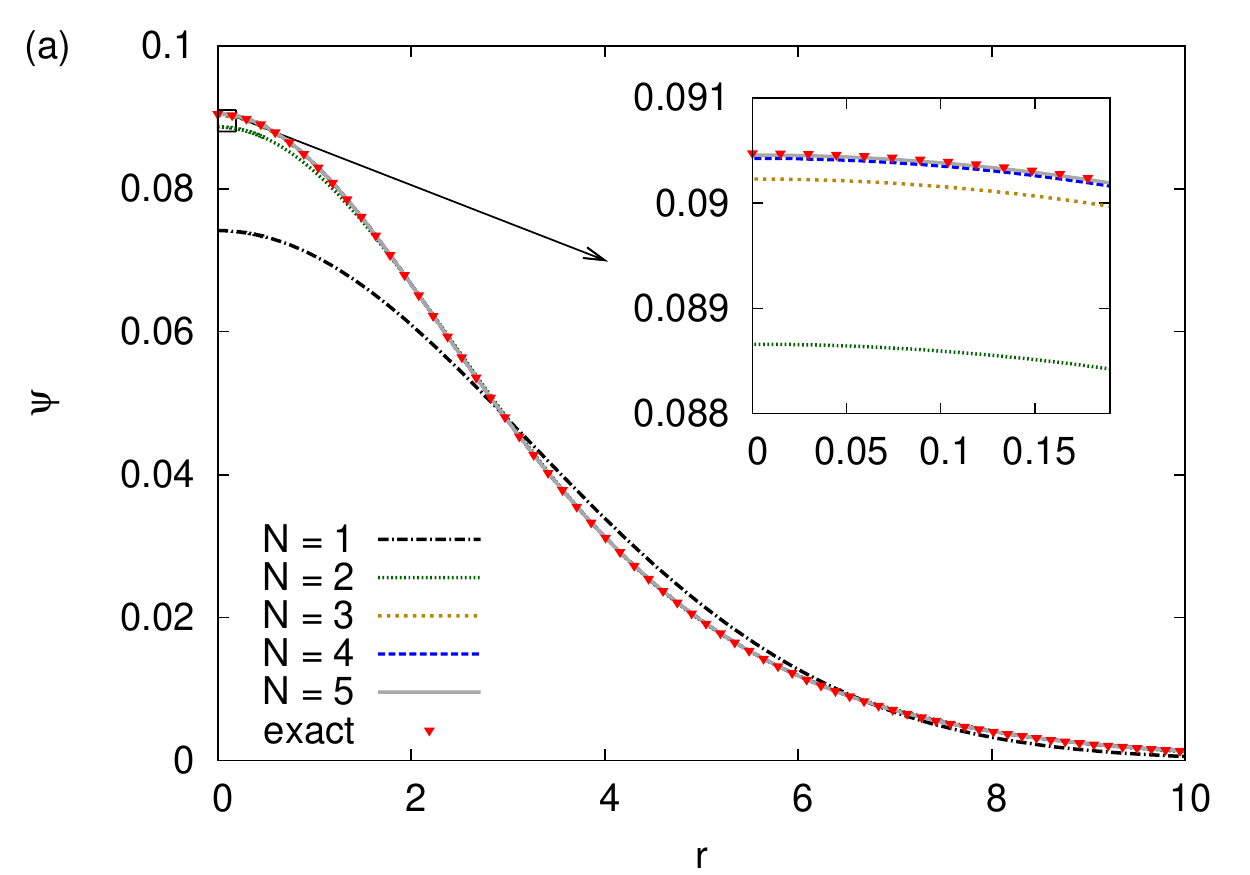}
\includegraphics[width=0.85\columnwidth]{Figure3a.pdf}
\caption{\label{fig:Psi_12345_b=-06}
 Same as Fig.~\ref{fig:Psi_12345_b=-1} but at scattering length $\scatt=-0.6$.}
\end{figure}
In Fig.~\ref{fig:Psi_12345_b=-1} we choose the scattering length 
$\scatt=-1$ close to the exact numerical bifurcation at 
$\scatt^\mathrm{cr} = -1.02515$ to compare the wave function of the 
variational and the numerically exact ground and excited state.
As a second example, we choose $\scatt=-0.6$ in Fig.~\ref{fig:Psi_12345_b=-06},
which lies farther away from the critical scattering length. 
The rapid convergence of the variational wave functions with the number
of Gaussians increasing from $N=1$ to 5 is impressive as can be seen in
the insets in Figs.~\ref{fig:Psi_12345_b=-1} and \ref{fig:Psi_12345_b=-06}.
Note that far from the bifurcation the wave function of the stable state 
in Fig.~\ref{fig:Psi_12345_b=-06}(a) and the wave function of the excited 
state in Fig.~\ref{fig:Psi_12345_b=-06}(b) differ greatly, while for 
$\scatt = - 1$ both wave functions in Fig.~\ref{fig:Psi_12345_b=-1}(a) 
and (b) are similar.
At the tangent bifurcation, both merging states, the stable ground state
and the excited state are described by the same wave function. 
The reason is that due to the nonlinearity of the GPE, the solution at the 
tangent bifurcation has the properties of an ``exceptional point'' 
\cite{holgerPRA77}.

\begin{figure}
\includegraphics[width=0.85\columnwidth]{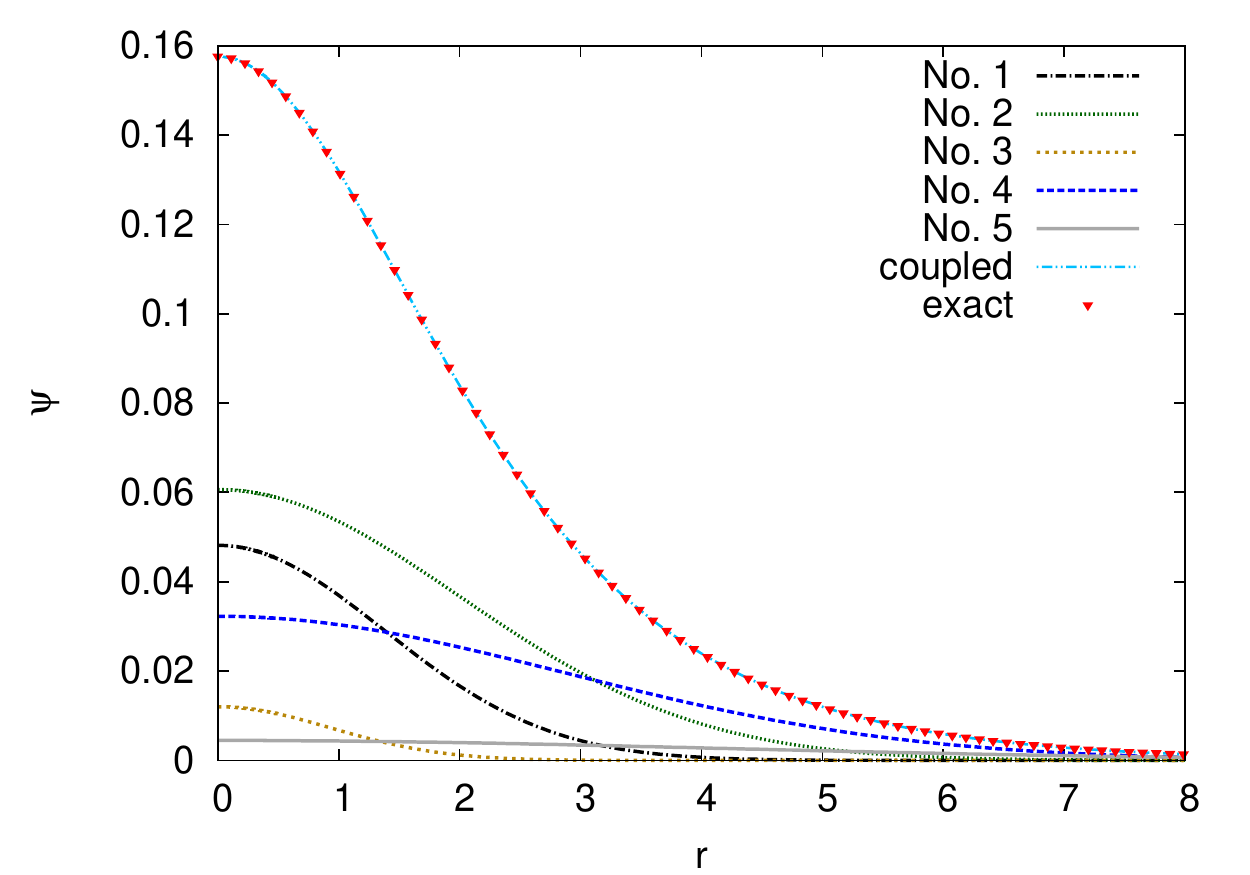}
\caption{\label{fig:psiaddition}
 The constituents of the five-Gaussian ansatz (No. 1, $\hdots$, 5) plotted
 separately.  The superposition of the five Gaussians (coupled), excellently
 agrees with the exact numerical solution (exact) at scattering length
 $\scatt = -1$.}
\end{figure}
Both Figs.~\ref{fig:Psi_12345_b=-1} and \ref{fig:Psi_12345_b=-06} apparently 
show that the form of the exact numerical wave function differs from the 
simple $N=1$ Gaussian form. 
The inclusion of five Gaussians, however, achieves excellent results.
Therefore we can assume the variational ansatz using five coupled Gaussian 
functions to be converged with sufficient accuracy to calculate all major 
properties of the system.

To visualize the ansatz, the five Gaussian functions as constituents 
of the $N=5$ solution at scattering length $\scatt = -1$ in 
Fig.~\ref{fig:Psi_12345_b=-1} are drawn in Fig.~\ref{fig:psiaddition}.
The two nearly identical curves at the top are the variational solution 
given as the sum of the functions below, and, for comparison, the exact 
numerical solution (triangles).

\subsection{Stability of stationary states}
\label{subsec:stabilitycoupled_monopolar}
For the spherically symmetric monopolar condensate we restrict the 
discussion of the stability of states to fluctuations of the wave function
in the radial coordinate $r$. 
In numerically exact calculations, the linearization of the GPE with the 
Fr\'echet derivative \cite{holgerPRA78} leads to two coupled Bogoliubov
equations for the real and imaginary parts of the wave function 
$\psi^\mathrm{Re}(r,t)$ and $\psi^\mathrm{Im}(r,t)$,
\begin{subequations}
\begin{align}
 \deriv{}{t} \delta\psi^\mathrm{Re}(r,t) &= 
 \Bigg( -\Delta - \mu + 8 \pi \scatt \hat \psi_{\pm}(r)^2 \nonumber \\
 &- 2 \int \mathrm{d}^3\bm r' \frac{\hat \psi_{\pm}(r')^2 }
 {\left| \bm r - \bm r' \right|}  \Bigg)\delta\psi^\mathrm{Im}(r,t) \; ,\\
\deriv{}{t} \delta\psi^\mathrm{Im}(r,t) &= 
\Bigg( -\Delta - \mu + 24 \pi \scatt \hat \psi_{\pm}(r)^2 \nonumber \\
 &- 2 \int \mathrm{d}^3\bm r' \frac{\hat \psi_{\pm}(r')^2 }
 {\left| \bm r - \bm r' \right|} \Bigg)\delta\psi^\mathrm{Re}(r,t) \nonumber \\
 &+4 \hat \psi_{\pm}(r)
\int \mathrm{d}^3\bm r' \frac{\hat \psi_{\pm}(r')
  \delta \psi^{\mathrm{Re}}(r',t) }{\left| \bm r - \bm r' \right|} \; ,
\end{align}
\label{eq:bogo}
\end{subequations}
where $\hat \psi_{\pm}(r)$ is the numerically exact stationary ground or 
excited state, respectively.
The method for solving those equations with the ansatz for the perturbations 
\begin{subequations}
\begin{align}
 \delta\psi^\mathrm{Re}(r,t) &= \delta\psi^\mathrm{Re}_0(r) \ee^{\lambda t} \; ,\\
 \delta\psi^\mathrm{Im}(r,t) &= \delta\psi^\mathrm{Im}_0(r) \ee^{\lambda t} \; ,
\end{align}
\label{eq:Bogo_ansatz}
\end{subequations}
where $\lambda$ is one of the exact stability eigenvalues, is elaborated
in \cite{holgerPRA78}.
The exact stability eigenvalues are used for comparisons with the
eigenvalues of the Jacobian
\begin{equation}
\label{eq:theory_Jacobian_EW}
 \bm{J} = \frac{ \partial \left(\gamma^{k,\mathrm{Re}},\gamma^{k,\mathrm{Im}}
    , a^{k,\mathrm{Re}}, a^{k,\mathrm{Im}} \right)}
		{ \partial \left(\gamma^{l,\mathrm{Re}},\gamma^{l,\mathrm{Im}}
    , a^{l,\mathrm{Re}}, a^{l,\mathrm{Im}} \right)} \; ; \; k,l=1, \dots, N ,
\end{equation}
which is a $(4N\times 4N)$-dimensional non-symmetric real matrix obtained 
by linearization of the dynamical equations of motion 
\eqref{eq:eomReducedSymmetry_r} for the variational parameters \cite{paper1}.
The eigenvalues of the Bogoliubov equations \eqref{eq:bogo} with the ansatz
\eqref{eq:Bogo_ansatz} and of the Jacobian $\bm{J}$ in 
Eq.~\eqref{eq:theory_Jacobian_EW} always occur in pairs $\pm\lambda$ 
with opposite sign.

\subsubsection{First pair of stability eigenvalues} 
\begin{figure}
\includegraphics[angle=-90, width=0.9\columnwidth]{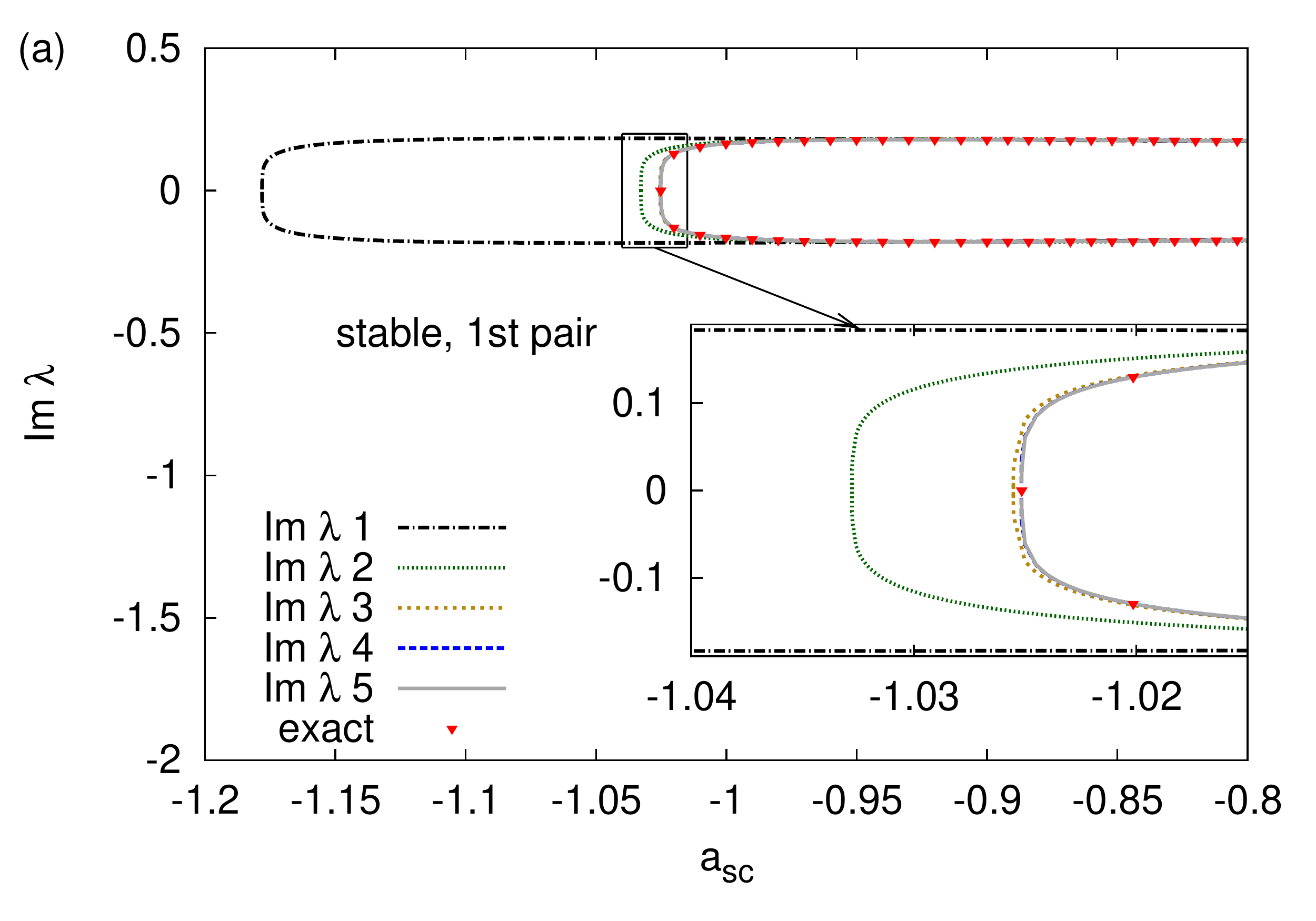}
\includegraphics[angle=-90, width=0.9\columnwidth]{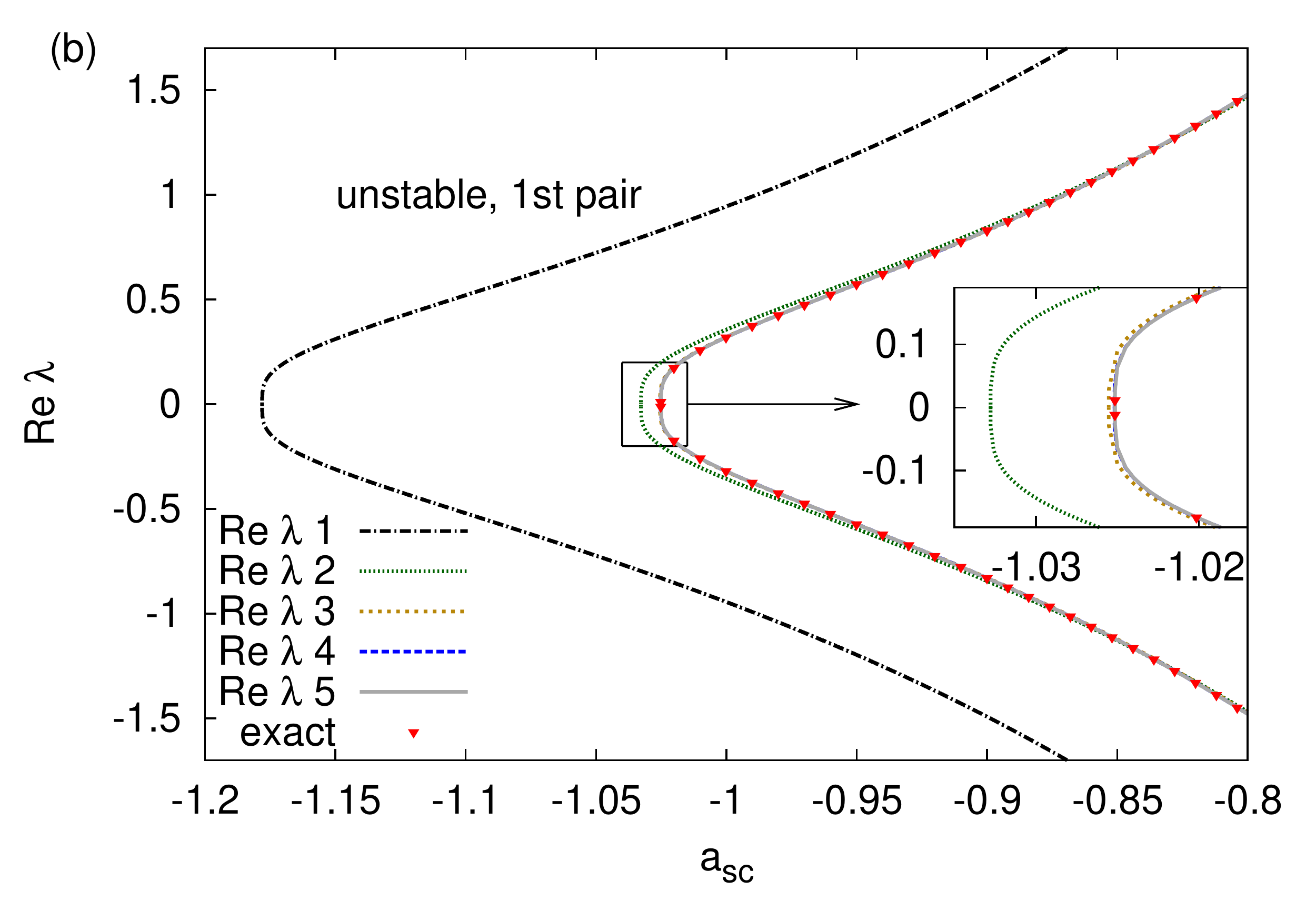}
\caption{\label{fig:EWcoupledlowest}
 First pair of eigenvalues as a function of the scattering length, for an 
 increasing number of coupled Gaussians and for the exact numerical
 calculation. 
 (a) The two lowest eigenvalues of the stable ground state are purely imaginary.
 (b) The unstable excited state with a pair of purely real eigenvalues. 
 Vanishing real or imaginary parts are not shown. 
 The insets illustrate the rapid convergence of the variationally computed
 eigenvalues to the exact solutions of the Bogoliubov equations.
}
\end{figure}

For the ansatz with a {\em single} Gaussian there is, after exploiting 
the normalization condition, only one variational parameter $a=a^1$ in 
Eq.~\eqref{eq:coupledAnsatz} for the (complex) width of the Gaussian 
function, and the eigenvalues of the Jacobian can be calculated 
analytically \cite{holgerPRA78,Gio01}.
For the ground state the two eigenvalues
\begin{equation}
 \lambda^\mathrm{g}_\pm = \pm \frac{16 \ii}{9 \pi}
 \frac{\sqrt[4]{1 + \frac{8 \scatt}{3 \pi} }}
 {\left(\sqrt{1+ \frac{8 \scatt}{3 \pi}}+1\right)^2} \; ,
\label{eq:lambda_g}
\end{equation}
are purely imaginary for all scattering lengths above the critical value
of the tangent bifurcation.
For the excited state there are two purely real eigenvalues
\begin{equation}
 \lambda^\mathrm{e}_\pm = \pm \frac{16}{9 \pi}
\frac{\sqrt[4]{1 + \frac{8 \scatt}{3 \pi} }}
{\left(\sqrt{1+ \frac{8 \scatt}{3 \pi}}-1\right)^2} \; .
\label{eq:lambda_e}
\end{equation}
For $N\ge 2$ coupled Gaussian functions the Jacobian $\bm J$ is diagonalized 
numerically.
The stability eigenvalues obtained with a single Gaussian function
(Eqs.~\eqref{eq:lambda_g} and \eqref{eq:lambda_e}),
the corresponding pair of eigenvalues of the Jacobian $\bm J$ for $N=2$ 
to $N=5$ coupled Gaussians, and the corresponding pair of the exact 
eigenvalues are presented in Fig.~\ref{fig:EWcoupledlowest}.
Similar to the calculation of the mean field energy and chemical potential,
the pair of eigenvalues merges and vanishes at a tangent bifurcation. 
As we include more Gaussian functions, the critical scattering length shifts 
to higher values and converges to the bifurcation of the exact numerical 
solution as was expected from the behavior of the energies.
Fig.~\ref{fig:EWcoupledlowest}(a) shows the first pair of eigenvalues of 
the stable ground state. 
They are purely imaginary.
Since this is true for all stability eigenvalues of the ground state (see below),
the branch of the ground state is stable. 
Fig.~\ref{fig:EWcoupledlowest}(b) shows the first eigenvalue pair of the
excited state.
These eigenvalues are purely real, and thus the excited state is unstable.
The tangent bifurcation is clearly exhibited for both states, as each pair 
of eigenvalues $\pm\lambda$ merges at zero and vanishes at the critical 
scattering length.

It is important to note that similar to the convergence properties of the 
mean field energy and the chemical potential, very few coupled Gaussians 
are already sufficient to achieve excellent results for the lowest stability 
eigenvalues. 
Results obtained with $N\ge 3$ coupled Gaussians can not be distinguished 
in Fig.~\ref{fig:EWcoupledlowest} from the numerically exact values.

\subsubsection{Additional pairs of stability eigenvalues}
\begin{figure}
\includegraphics[angle=-90, width=0.85\columnwidth]{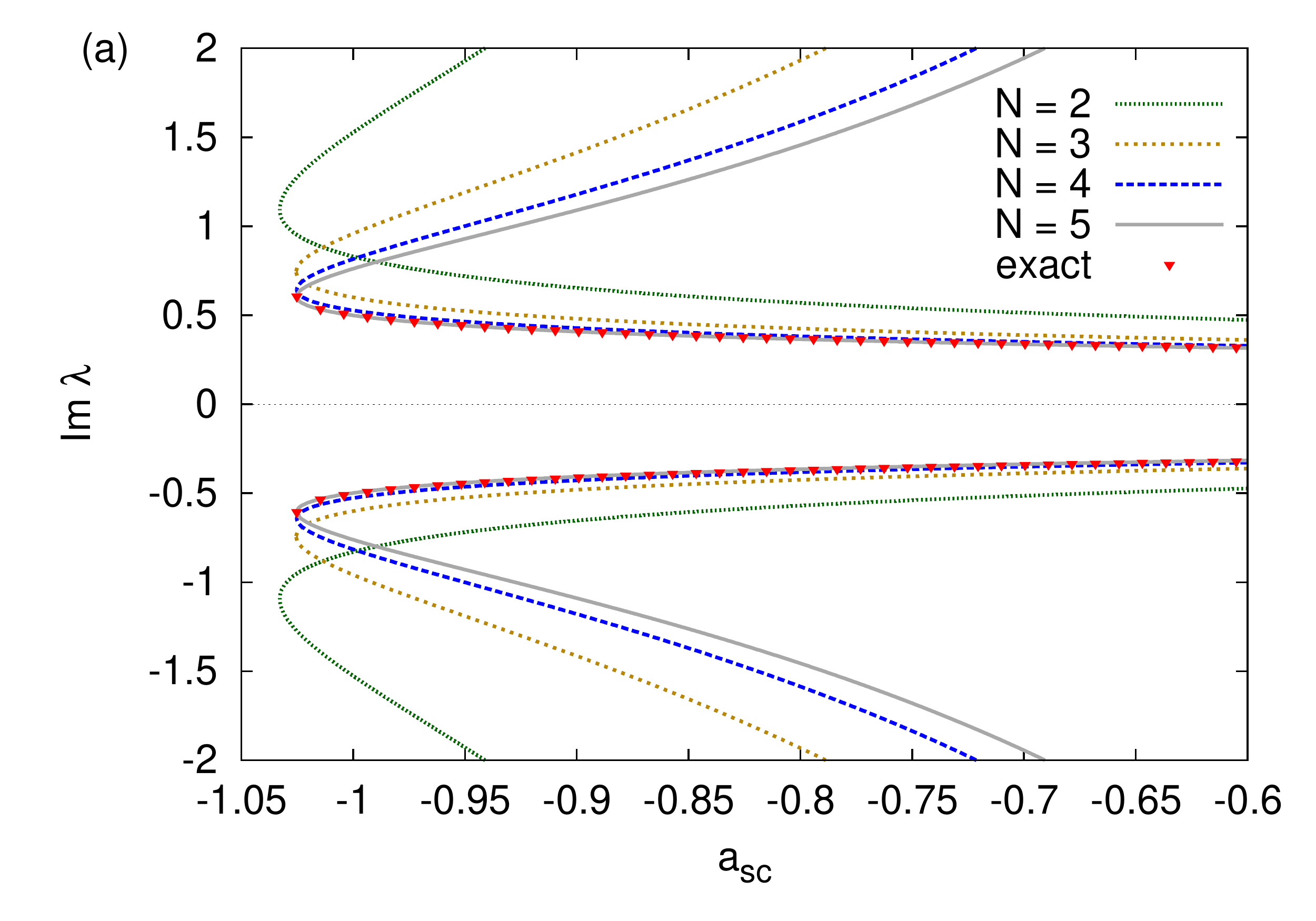}
\includegraphics[angle=-90, width=0.85\columnwidth]{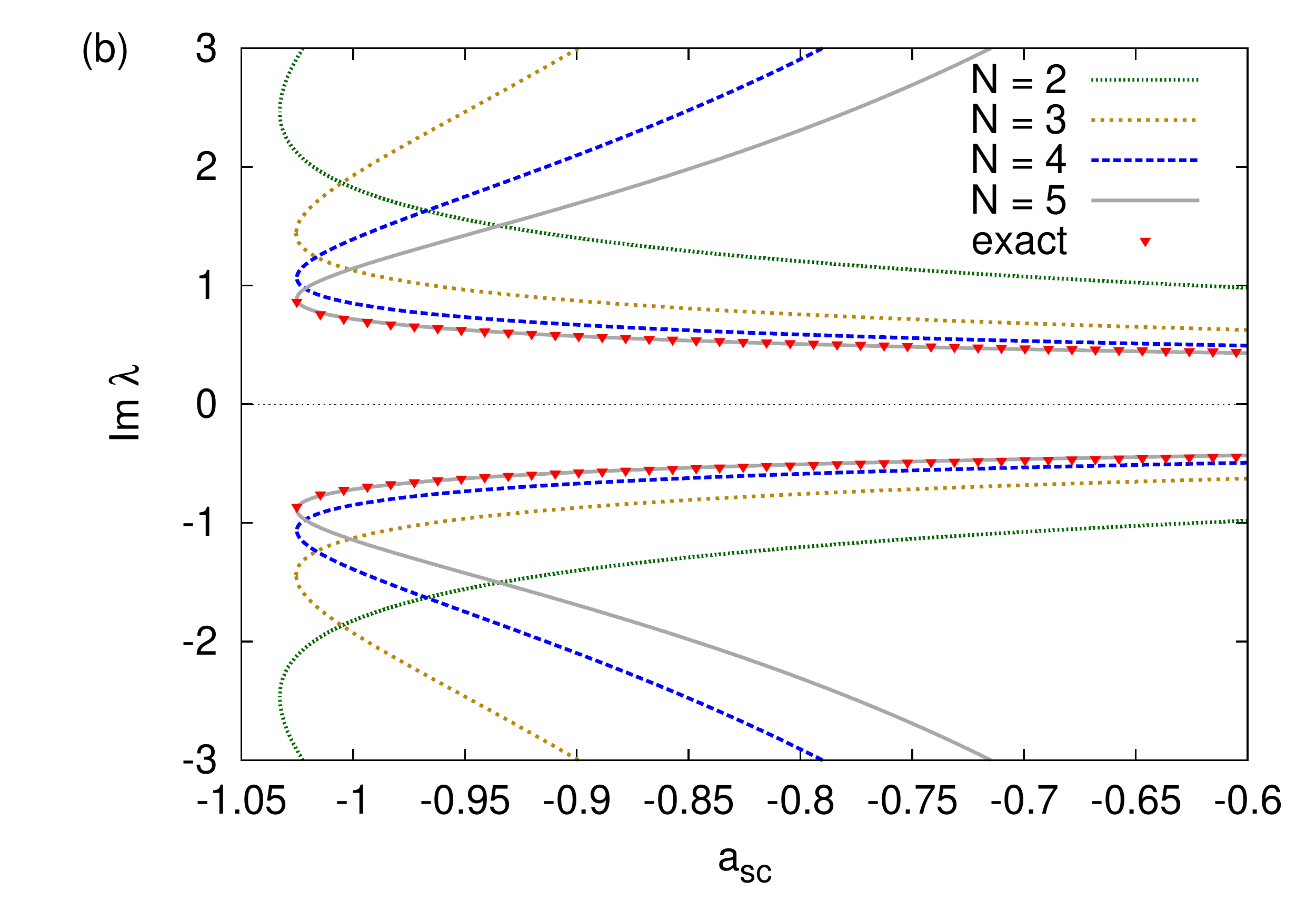}
\caption{\label{fig:EWcoupled2ndand3rdlowest}
 (a) Second pair of eigenvalues as a function of the scattering length, for
 varying number of coupled Gaussians and for the exact numerical calculation 
 for the stable ground state and the unstable excited state. 
 Numerically exact eigenvalues (triangles) are only available for the stable
 ground state.  All eigenvalues are purely imaginary, vanishing real parts
 are not shown.  (b) Same as (a) but for the third pair of eigenvalues.}
\end{figure}

In contrast to the calculation with a single Gaussian, which can provide 
only one pair of eigenvalues, additional eigenvalues are accessible when 
using coupled Gaussian wave functions.
We compare them with the exact stability eigenvalues 
in Fig.~\ref{fig:EWcoupled2ndand3rdlowest}.
The additional eigenvalues are increasingly difficult to obtain in the 
numerically exact computation.
In the following, we refer to the eigenvalues with the second lowest absolute 
value simply as ``second (pair of)'' eigenvalues, etc.
Numerically exact data is available for the lowest three pairs of eigenvalues 
of the stable ground state, and therefore here we compare the variational 
solution only with the lowest three eigenvalues of the stable solution.

Figure \ref{fig:EWcoupled2ndand3rdlowest}(a) shows the second pair of 
eigenvalues for the stable ground state and the unstable excited state.
The second pairs of eigenvalues are all purely imaginary, and converge with 
increasing number of coupled Gaussians to the numerically exact eigenvalues.

The third pair of eigenvalues presented 
in Fig.~\ref{fig:EWcoupled2ndand3rdlowest}(b) is qualitatively similar to 
the second pair.
For any number of coupled Gaussians, they are purely imaginary.
Figs.~\ref{fig:EWcoupled2ndand3rdlowest}(a) and (b) suggest that, as the 
number of the eigenvalue rises, the convergence progresses more slowly. 
However, using five Gaussians in Fig.~\ref{fig:EWcoupled2ndand3rdlowest}(b) 
even the third pair of the variational eigenvalues shows no apparent 
deviation from the numerically exact result.

\subsubsection{Variations of the ground state wave function}
\label{par:monopolardeltapsi}
We investigated the stability of the stationary states by linearizing 
the dynamical equations in the vicinity of the fixed points.
For the stable ground state there are only purely imaginary eigenvalues,
for the unstable excited state we found one pair of real eigenvalues. 
Additionally to the analysis of the eigenvalues $\lambda_i$ of the Jacobian 
of the linearized dynamical equations, we can evaluate the respective 
eigenvectors, which provide the form of the wave function's fluctuations.

\begin{figure}
\includegraphics[width=0.85\columnwidth]{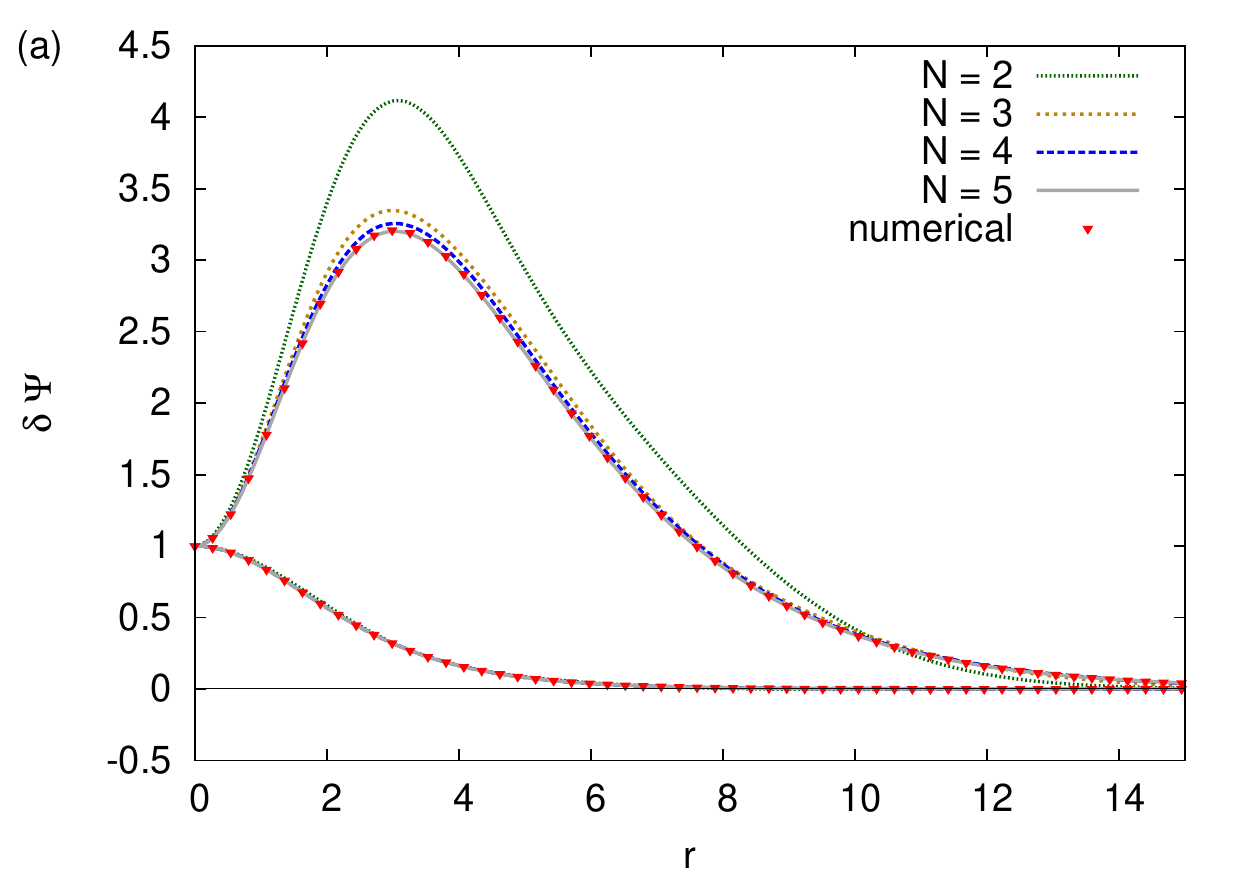}
\includegraphics[width=0.85\columnwidth]{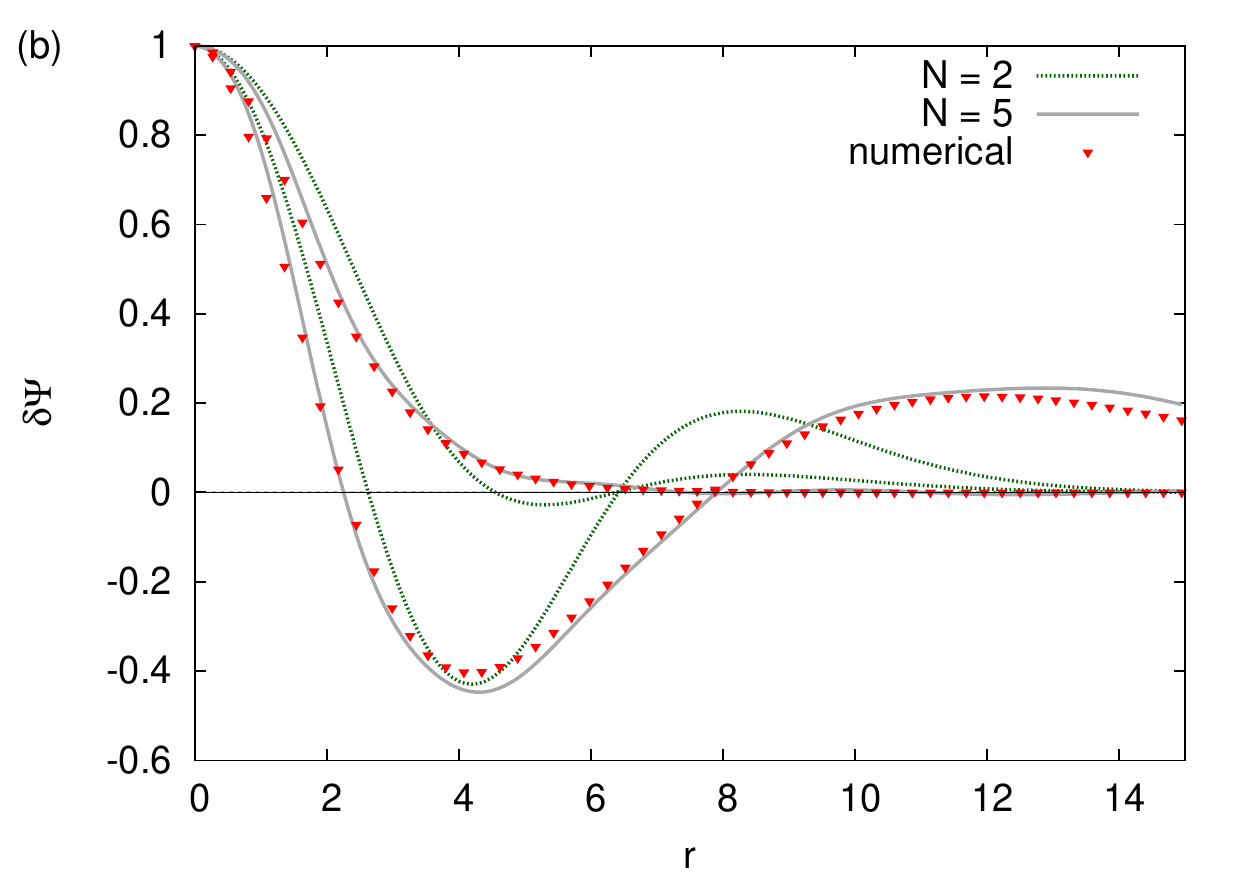}
\includegraphics[width=0.85\columnwidth]{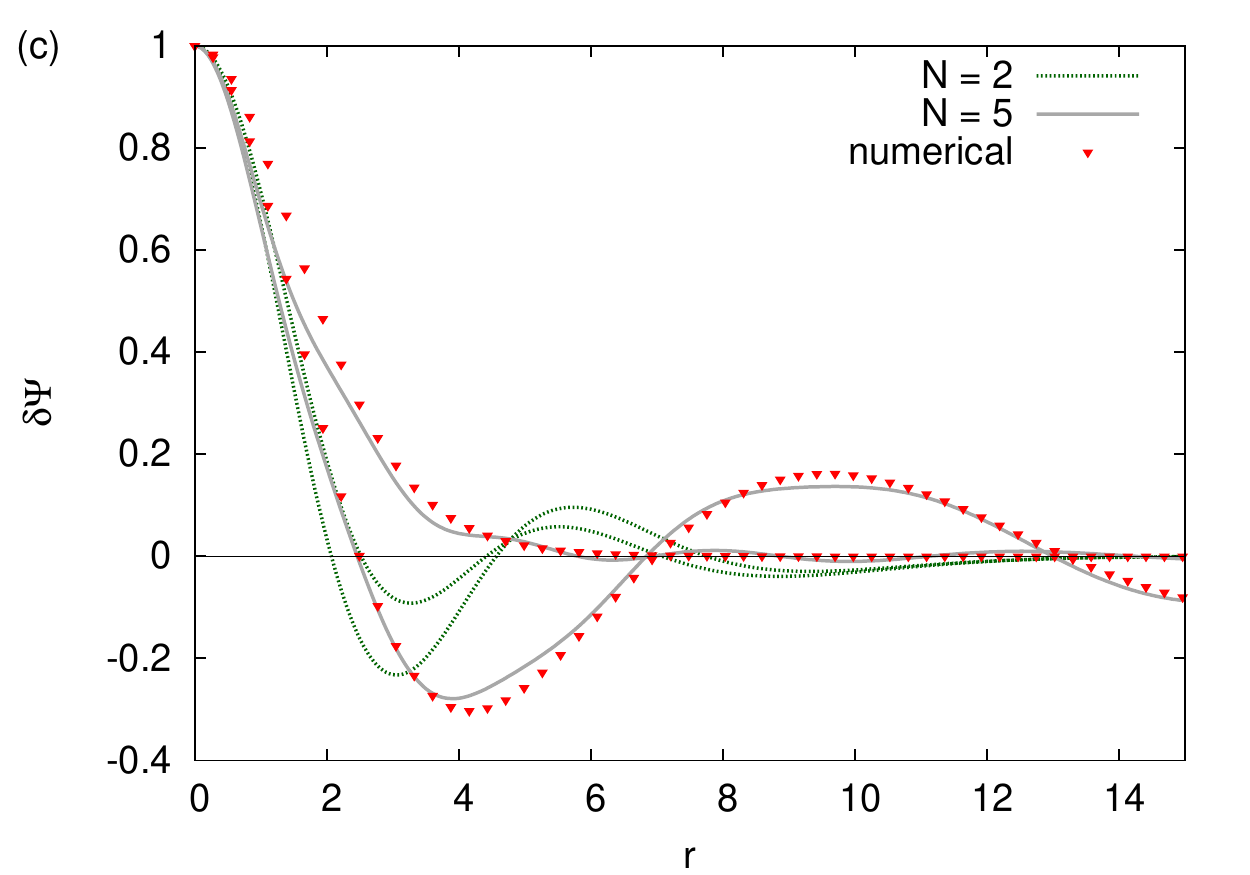}
\caption{\label{fig:psideviation12}
 Deviations of the ground state wave function according to the eigenvectors
 of the (a) first, (b) second, and (c) third pair of eigenvalues $\pm\lambda$
 of the stability analysis at scattering length $\scatt = -0.8$.}
\end{figure}

We focus on variations of the Gaussian parameters 
corresponding to the eigenvector $i$ and calculate the first order power
series of the wave function $\psi(r,t)$ at the fixed point (FP) \cite{paper1},
\begin{align}
\delta \psi_i(r,t) = \sum_{k = 1}^{N}
 &\left( \ii r^2 \delta a_i^{k,\mathrm{Re}} - r^2 \delta a_i^{k,\mathrm{Im}} + \ii \delta \gamma_i^{k,\mathrm{Re}} - \delta \gamma_i^{k,\mathrm{Im}}\right) \nonumber \\
&\times g^k|^{\mathrm{FP}}(r)
\, \ee^{\lambda_i t}.
\end{align}
For a scattering length of $\scatt=-0.8$, Fig.~\ref{fig:psideviation12}
shows the variation of the ground state wave function for the eigenvectors 
corresponding to the first (a), second (b) and third (c) pair of eigenvalues 
$\pm\lambda$ for the variational calculation, as well as for the numerically 
exact calculation.

In Fig.~\ref{fig:psideviation12}(a) the variation of the wave function 
$\delta \psi$ converges rapidly with increasing number of Gaussian functions 
to the numerical solution. 
For the second and third pair of eigenvalues in Fig.~\ref{fig:psideviation12}(b)
and (c), the variation using five Gaussians is almost identical to the 
numerical variation.
For the higher pairs of eigenvalues, the solution obviously converges slower.
However, we note that it also becomes more and more difficult to obtain the
exact solutions.
The spatial extension of the fluctuations $\delta\psi$ exceeds the elongation
of the wave function $\psi$ (cf.\ Figs.\ \ref{fig:Psi_12345_b=-1} and 
\ref{fig:Psi_12345_b=-06}) by far and it becomes hard to achieve converged
fluctuations.
The differences between the numerical and the $N=5$ solution in 
Fig.~\ref{fig:psideviation12}(c) can already be explained by the quality
of the numerical approach.

\section{Dipolar condensates}	
\label{chap:dipolar_interaction}
The stationary GPE for a dipolar BEC with a harmonic trap, the short-range 
s-wave scattering term and the long-range dipolar interaction reads
\begin{align}
 \bigg[ &- \Delta 
   + \gamma_x^2 x^2 + \gamma_y^2 y^2 + \gamma_z^2 z^2 
   + 8 \pi N \frac{a}{a_d} \left |  \psi(\bm{r})\right |^2 \nonumber \\
   &+ N  \int \mathrm{d}^3 \bm{r}' \frac{1 - 3 \cos^2\theta} 
  {\left | \bm{r} - \bm{r}' \right |^3}\left | \psi(\bm{r}')\right |^2 \bigg]
  \psi(\bm r) = \mu \psi(\bm r) \; .
\label{eq:GPE_dimless_dipolar}
\end{align}
In Eq.~\eqref{eq:GPE_dimless_dipolar} the ``natural units'' for this system 
introduced in \cite{gelbPatrick} have been used, which are
$\hbar$ for action, $m_d=2m$ for mass, $a_d={m_d \mu_0 \mu^2}/(4\pi\hbar^2)$
for length, $E_d={\hbar^2}/(m_d a_d^2)$ for energy, and $\omega_d=E_d / \hbar$
for frequency.
The angle between the external magnetic field in $z$-direction and the 
vector $\bm{r}-\bm{r}'$ is denoted $\theta$, and $N$ is the particle number.
As in \cite{gelbPatrick} we scale ${\bm r}= N\tilde{\bm r}$,
$\psi = N^{-\nicefrac{3}{2}}\tilde \psi$,
insert the newly defined quantities in Eq.~\eqref{eq:GPE_dimless_dipolar}, 
redefine $\tilde\gamma_i = N^2 \gamma_i$, $\tilde\mu = N^2\mu$,
and afterwards omit the tilde once again. 
With the replacement $\mu\to \ii(\dd/\dd t)$ we finally obtain the 
time-dependent GPE for a dipolar BEC in particle number scaled 
dimensionless units,
\begin{align}
    \bigg[ &- \Delta 
       + \gamma_x^2 x^2 + \gamma_y^2 y^2 + \gamma_z^2 z^2 
             + 8 \pi  \scatt \left |  \psi(\bm{r})\right |^2 \nonumber \\
   &+   \int \mathrm{d}^3 \bm{r}' \frac{1 - 3 \cos^2\theta} 
  {\left | \bm{r} - \bm{r}' \right |^3}\left | \psi(\bm{r}')\right |^2 \bigg]
  \psi(\bm r) = \ii\frac{\dd}{\dd t} \psi(\bm r) \; ,
\label{eq:dipolarGPEmu1}
\end{align}
with the trap frequencies $\gamma_{x,y,z}=N^2\omega_{x,y,z}/(2\omega_d)$
and the s-wave scattering length $\scatt=a/a_d$.
For dipolar condensates it is possible to solve the GPE fully numerically
on a two- or three-dimensional lattice \cite{Lahaye2009,Ronen06a}.
Ronen et al.\ \cite{Ronen} and Dutta et al.\ \cite{duttadelle} have shown 
that in certain regions of the parameter space dipolar condensates assume 
a non-Gaussian, biconcave, ``blood-cell-like'' shape.
In this paper we want to apply the variational method with coupled Gaussian
functions introduced in the preceding paper \cite{paper1}.
We will show that the variational technique is a full-fledged alternative
to the numerical simulations on grids, and additionally uncovers unstable
stationary solutions not accessible in previous full-numerical evaluations.

The frequency and symmetry of the magnetic trap strongly influences the 
physical behavior of dipolar Bose-Einstein condensates.
In the following we will analyze one distinct trap symmetry in detail,
where the condensate has a blood-cell-shaped form.
The ansatz with coupled Gaussians has proved to be capable to modify the 
simple Gaussian form of the wave function for monopolar condensates 
(see Sec.~\ref{subsection:stationarySolutions_coupled_monopolar}). 
The biconcave shape, however, where the maximum density is no longer located 
at the origin is even a stronger challenge for the variational approach.
We investigate the dipolar condensate for an axially symmetric trap with 
trap frequencies 
\begin{align*}
 \gamma_x = \gamma_y \equiv \gamma_\varrho = 3600, \; \gamma_z=25200,
\end{align*}
which is equivalent to the frequency ratio and mean
\begin{align*}
 \lambda = \frac{\gamma_z}{\gamma_\varrho} = 7, \;
 \bar\gamma =\sqrt[3]{\gamma_\varrho^2 \gamma_z}= 6887,
\end{align*}
and corresponds to a value of $D=\sqrt{\gamma_\varrho}/2=30$ in \cite{Ronen}.
The trapping frequency in the $z$-direction parallel to the orientation of
the dipoles is seven times larger than in the plane perpendicular to that
direction, and for some parameters $\scatt$ the ground state of the 
condensate has a biconcave, blood-cell-shaped form \cite{Ronen}.
In contrast to monopolar condensates where the inclusion of additional 
Gaussian functions provides an improved numerical accuracy of the results, 
dipolar condensates offer a wealth of new phenomena with increasing number 
of coupled Gaussian functions as will be shown below.

\subsection{Variational calculations with one and two Gaussian functions}
\label{subsec:dresults:2Gn}
As ansatz for the variational calculations we use the wave function 
\begin{equation}
 \psi (\bm{r},t) = 
 \sum_{k=1}^N \ee^{\ii\left(a_x^{k}x^{2}+a_y^{k}y^{2}+a_z^{k}z^{2}+\gamma^{k} \right)}
 \equiv \sum_{k=1}^N g^{k} \; ,
\label{defwdh:dipolartrialfunction}
\end{equation}
where $N$ is the number of coupled Gaussians. 
For an axisymmetric trap the stationary solutions are also symmetric, i.e.,
$a_x^k=a_y^k\equiv a_\varrho^k$.
Nevertheless, all stability properties have been computed with the fully
three-dimensional ansatz.
The case of a single Gaussian function ($N=1$) has been discussed in
\cite{gelbPatrick}.
In this section we demonstrate that results especially for the mean field 
energy and chemical potential are already substantially improved with the 
use of only $N=2$ coupled Gaussians.
Results of variational calculations with a single Gaussian function and 
numerical simulations on grids are employed for comparison and discussion.

\subsubsection{Stationary solutions}
\label{subsubsec:dresults:2GnMeanField}
Using the variational ansatz \eqref{defwdh:dipolartrialfunction}
stationary solutions of the Gross-Pitaevskii equation for dipolar BEC
are obtained by a numerical root search for the fixed points of the 
dynamical equations for the variational parameters as described in detail 
in the preceding paper \cite{paper1}.

\begin{figure}
\includegraphics[width=0.9\columnwidth]{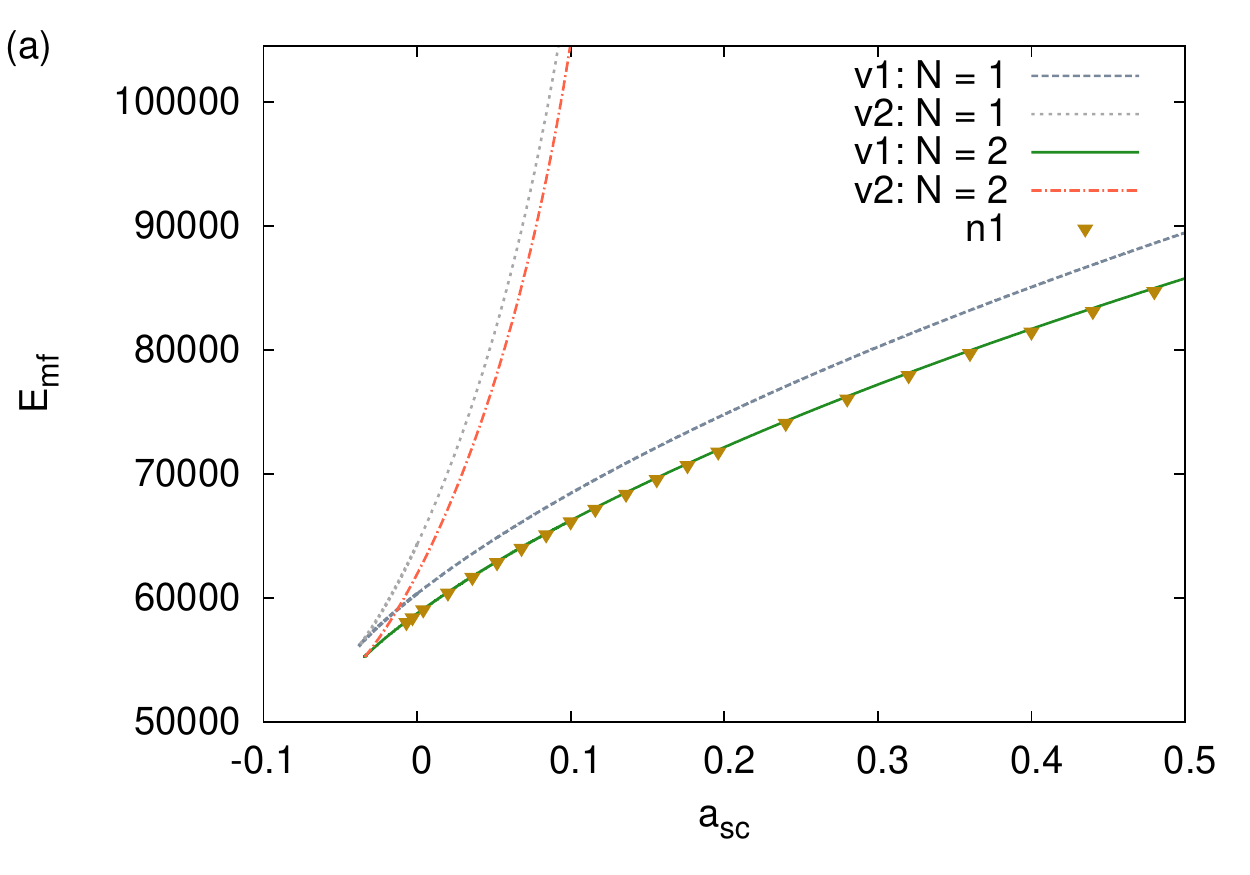}
\includegraphics[width=0.9\columnwidth]{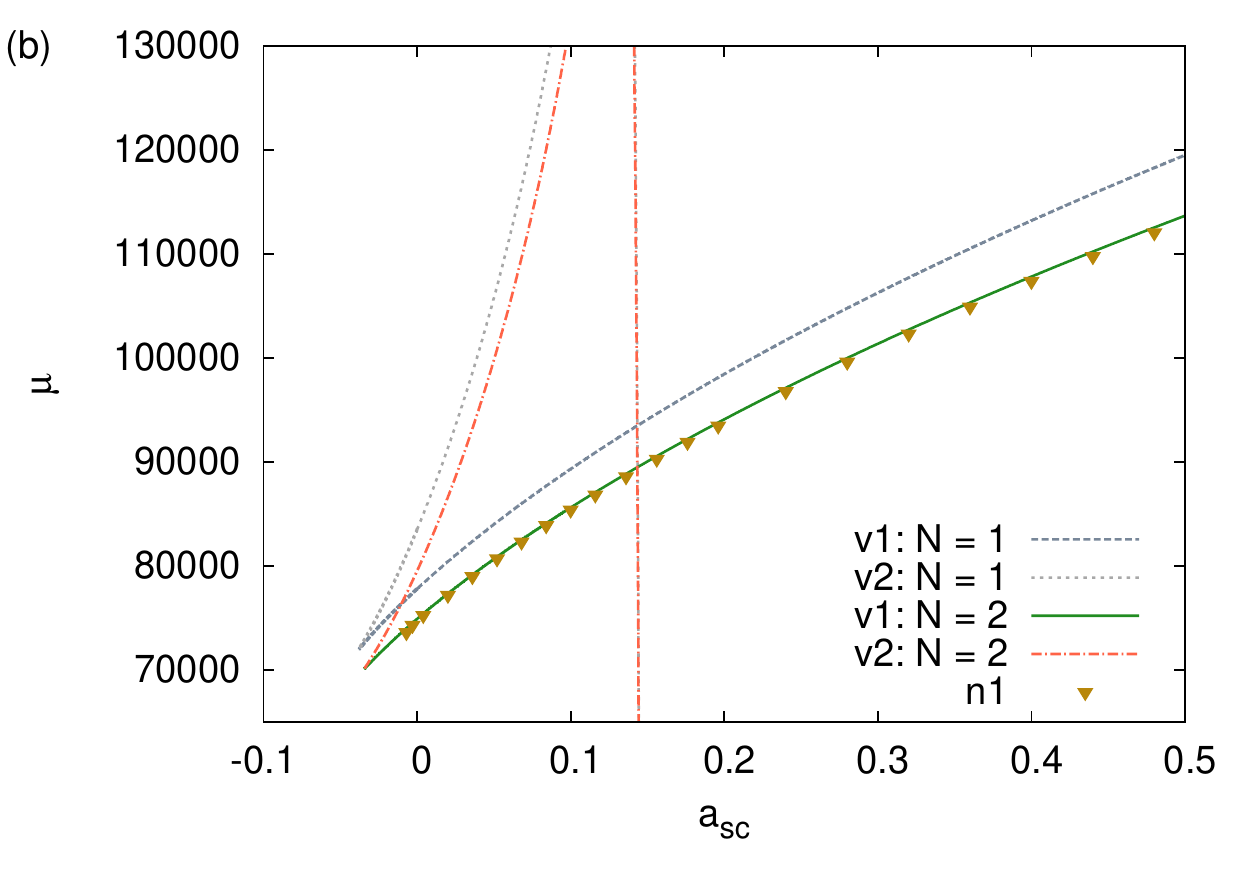}
\caption{\label{fig:E_2g_num_dipolar}
 (a) Mean field energy and (b) chemical potential as a function of the
 scattering length, variational solution (v) using two Gaussians compared
 with an exact full-numerical lattice calculation (n1).  
 For two Gaussians $(N = 2)$, the ground state (v1) and the excited
 state (v2) emerge in a tangent bifurcation at $\scatt = -0.034202$.
 The unstable branch of the chemical potential in (b) has a maximum at
 $\mu_{\max}\approx 170000$ (not shown) and then vanishes as a nearly vertical
 line at $\scatt=1/6$.}
\end{figure}

Figure~\ref{fig:E_2g_num_dipolar} 
presents in (a) the mean field energy and in (b) the chemical potential
for the two stationary solutions obtained with a single Gaussian function and
with two coupled Gaussians.
For the ground state results of a numerical lattice calculation 
are also marked in Fig.~\ref{fig:E_2g_num_dipolar}.
The numerical simulation was performed on a lattice with a grid size of 
$128 \times 512$ points using fast-Fourier techniques and imaginary time 
evolution of an initial wave function.

The variational calculations with {\em one} Gaussian ($N=1$) show the
following behavior.
For scattering lengths below $\scatt^{\mathrm{cr,var}}=-0.0378917$ there is 
no stable condensate.
Similar as in monopolar condensates two solutions are born at the critical 
scattering length in a tangent bifurcation, the stable ground state (v1) 
and an unstable excited state (v2). 
For a detailed stability analysis see Sec.\ \ref{subsubsec:dresults:1Gnstab} 
below. 
The unstable branch vanishes at scattering length $\scatt=1/6$.

The variational ansatz with $N=1$ is limited to the Gaussian shape of 
the wave function with two width parameters $a_\varrho$ and $a_z$, and
thus the values obtained for the mean field energy and chemical potential
are not very accurate.
However, the results are substantially improved when using a variational 
ansatz with {\em two} coupled Gaussians.
This can be seen in Fig.~\ref{fig:E_2g_num_dipolar} especially for the ground 
state when comparing the $N=2$ variational computation with the lattice 
computation (n1) marked by the triangles in Fig.~\ref{fig:E_2g_num_dipolar}.

In the full-numerical grid calculations only the ground state 
can be obtained. 
Starting with positive scattering lengths and decreasing $\scatt$, the 
numerical grid calculations provide a ground state down to a 
critical point $\scatt^{\mathrm{cr,num}}=-0.008$.
Note that the ground state of the solution using two coupled 
Gaussian wave functions is nearly indistinguishable from the numerical 
lattice calculation in Figs.~\ref{fig:E_2g_num_dipolar}(a) and (b). 
An important advantage of the variational method is that it can provide both 
stable and unstable states.
As will be shown below, the stability properties of the variational 
solutions can clarify the mechanism of how the condensate turns unstable.

Regarding the wave functions, one single Gaussian can evidently not 
adequately represent a blood-cell-shaped condensate. 
Two Gaussians not only significantly increase the accuracy of the mean field 
energy, but also greatly improve the form of the wave function.
Even the biconcave shape of the dipolar condensate is qualitatively visible 
in the variational solution with two-Gaussians, however, the result is
not fully converged.
Exact wave functions will be compared in Sec.~\ref{subsec:dresults:coupled}
with variational results obtained with more than two coupled Gaussians.

\subsubsection{Stability analysis}
\label{subsubsec:dresults:1Gnstab}
To perform a stability analysis with numerical lattice calculations 
the Bogoliubov-de Gennes equations have to be solved 
\cite{PitaevskiiBlackBook,Ronen06a}.
Here we restrict our discussion to the stability analysis of the 
variational solutions, which is instructive considering nonlinear 
dynamics and bifurcation theory. 

We follow the procedure outlined in \cite{paper1} and start with the 
stationary solutions of the GPE calculated using one and two Gaussians. 
These solutions are fixed points of the dynamical equations for the Gaussian 
parameters.
We then linearize these dynamical equations in the vicinity of the fixed 
point and calculate the eigenvalues of the Jacobian (see \cite{paper1}).
\begin{figure}
\includegraphics[angle=-90, width=0.85\columnwidth]{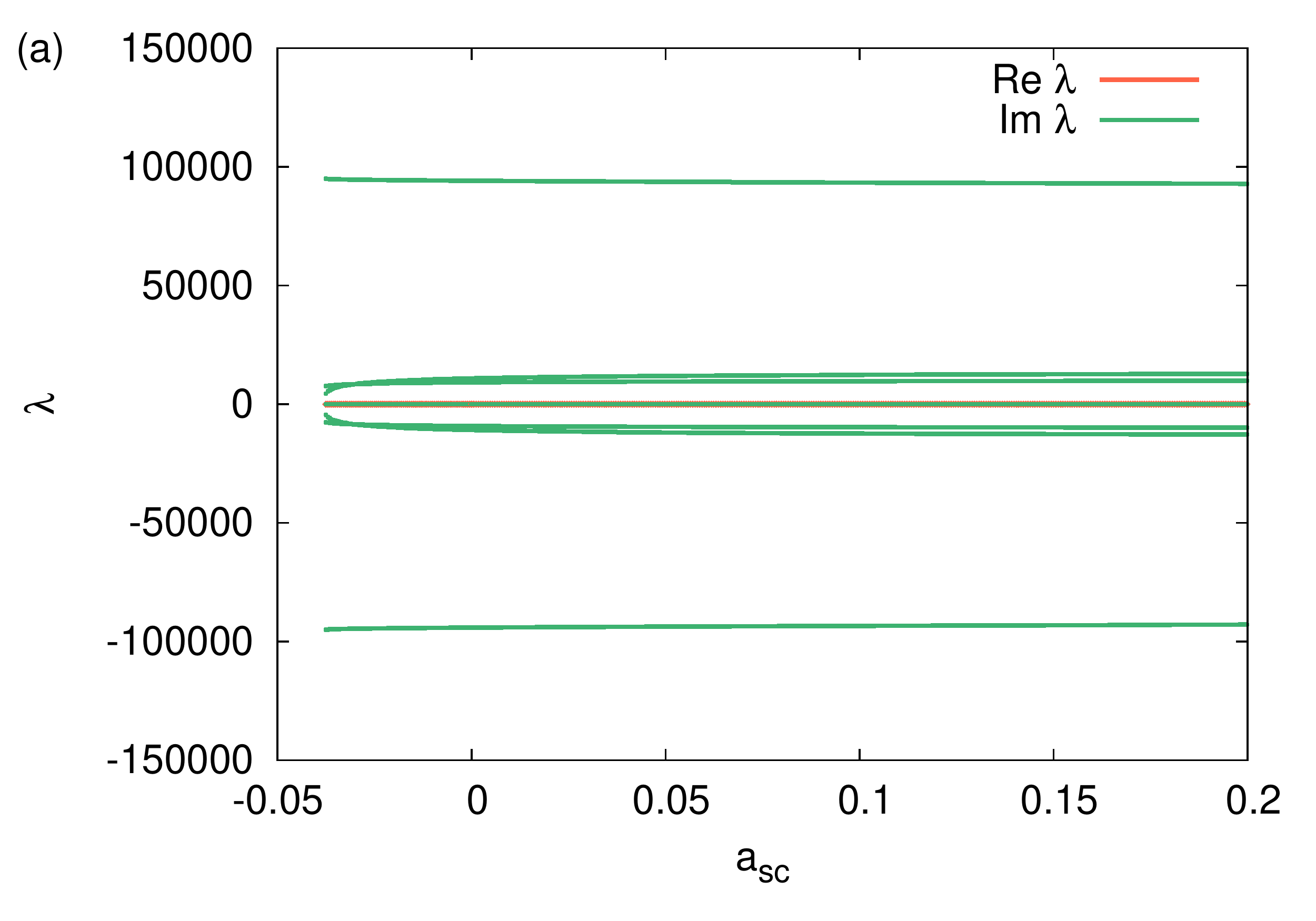}
\includegraphics[angle=-90, width=0.85\columnwidth]{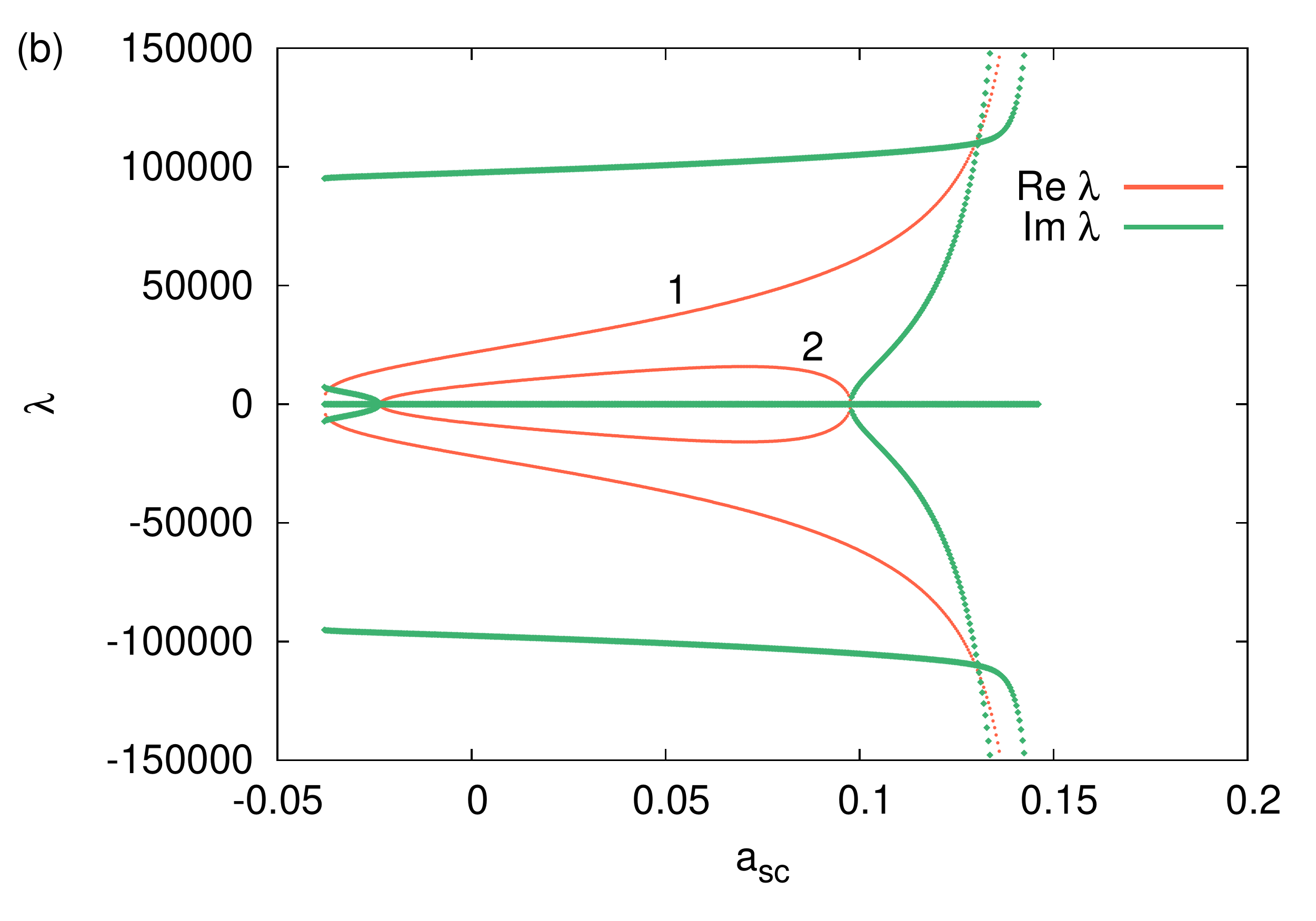}
\caption{\label{fig:EV1gn_dipolar}
 Eigenvalues of the Jacobian for the variational solution with one Gaussian
 as a function of the scattering length $\scatt$. 
 (a) The ground state is stable, all eigenvalues are purely imaginary,
 (b) the excited state is unstable since there are real parts of 
 eigenvalues, emerging in a tangent bifurcation at
 $\scatt^{\mathrm{cr}}=-0.0378917$.
 Eigenvalues which do not reach $\lambda=0$ at $\scatt^{\mathrm{cr}}$ match
 with the corresponding eigenvalues of the stable and unstable state,
 respectively.}
\end{figure}

The eigenvalues of the ground state obtained with {\em one} Gaussian wave 
function in Fig.~\ref{fig:EV1gn_dipolar}(a) are purely imaginary. 
Therefore this state is stable. 
If we perturb the variational parameters of the fixed point solution, 
the quasi-periodic motion is confined to the vicinity of the fixed point. 
\begin{figure}
\includegraphics[angle=-90, width=0.85\columnwidth]{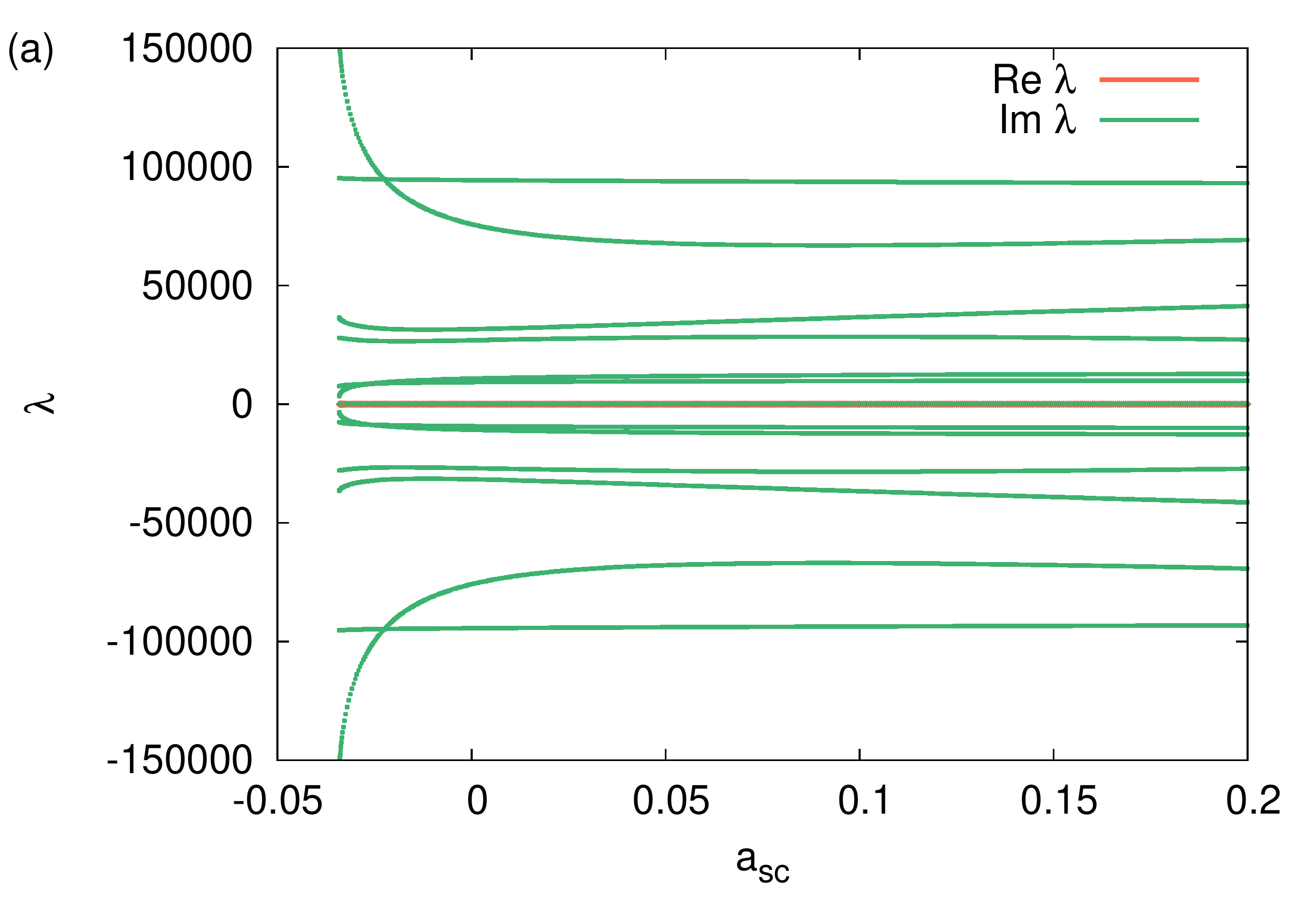}
\includegraphics[angle=-90, width=0.85\columnwidth]{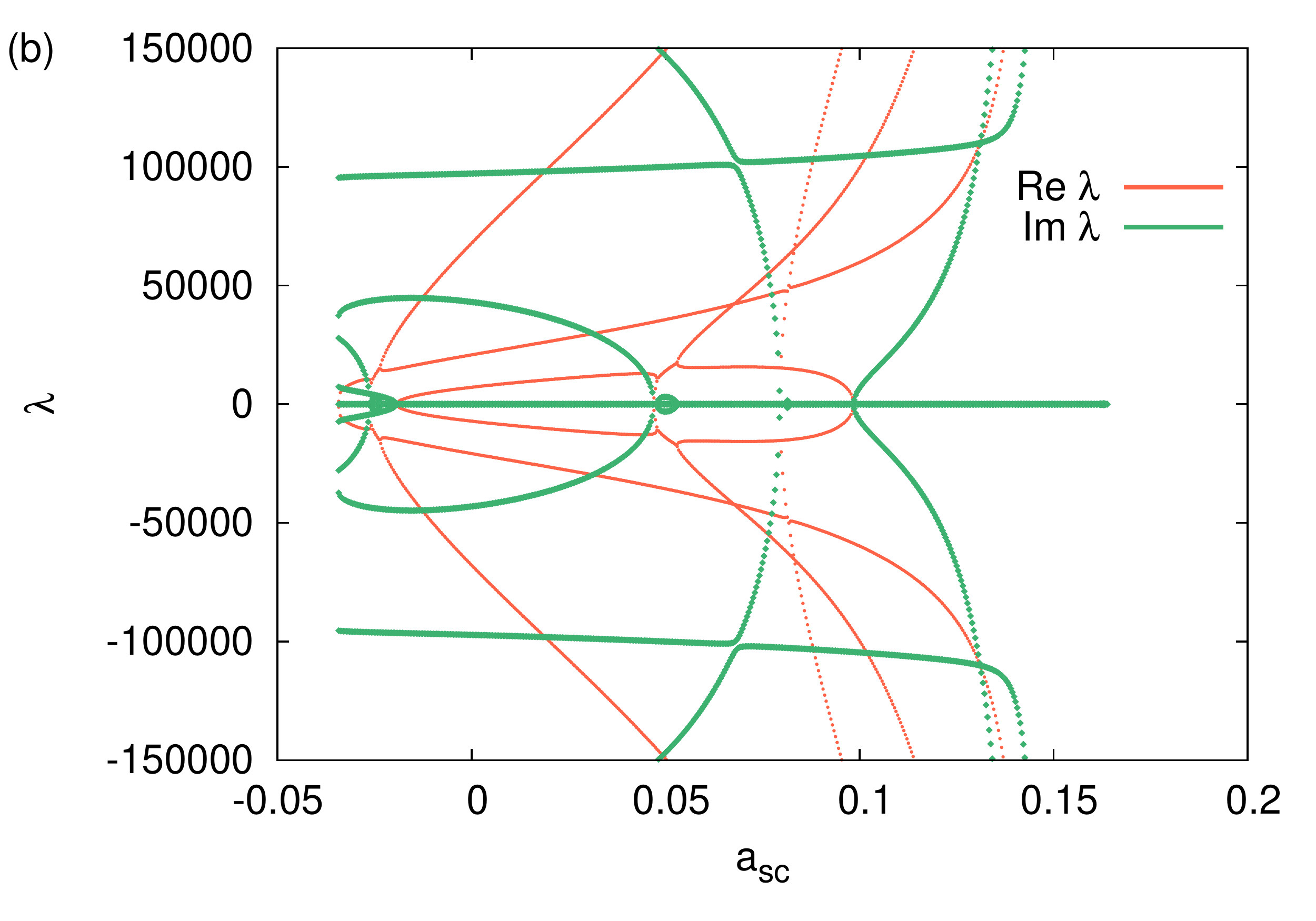}
\caption{\label{fig:lambda_2g_dipolar}
 Same as Fig.~\ref{fig:EV1gn_dipolar} but for the variational calculation
 with two coupled Gaussians, (a) the stable ground state and (b) the unstable
 excited state.  Additional eigenvalues are revealed.  The tangent bifurcation
 is shifted to $\scatt^{\mathrm{cr}}=-0.034202$.
 Eigenvalues which do not reach $\lambda=0$ at $\scatt^{\mathrm{cr}}$ match
 with the corresponding eigenvalues of the stable and unstable state,
 respectively.}
\end{figure}
Figure~\ref{fig:EV1gn_dipolar}(b) shows the characteristic eigenvalues of 
the unstable excited state. 
Contrary to the stable state, there are eigenvalues with non-vanishing real 
parts $\mathrm{Re}\,\lambda$.
Perturbations in the direction of the corresponding eigenvector lead to 
an exponential growth of the perturbation.
Therefore this state is unstable. 
Both branches exist for scattering lengths down to 
$\scatt^{\mathrm{cr}} = -0.0378917$, where they merge in a tangent bifurcation
(see Fig.~\ref{fig:E_2g_num_dipolar}).
The tangent bifurcation is apparent in the eigenvalues in 
Figs.~\ref{fig:EV1gn_dipolar}(a) and (b).

Some eigenvalues of the Jacobian using a wave function of {\em two} coupled 
Gaussians in Fig.~\ref{fig:lambda_2g_dipolar} qualitatively agree with the 
one-Gaussian calculation.
There is one stable ground state with purely imaginary eigenvalues 
in Fig.~\ref{fig:lambda_2g_dipolar}(a), and one unstable excited state 
in Fig.~\ref{fig:lambda_2g_dipolar}(b). 
However, the two-Gaussian calculation yields additional eigenvalues which 
are not available within the limited parameter space of the simple 
one-Gaussian calculation.

The unstable branches of both calculations in Figs.~\ref{fig:EV1gn_dipolar}(b) 
and \ref{fig:lambda_2g_dipolar}(b), exhibit that for the given 
parameters of the trap in dipolar condensates, the stability scenario is 
quite more complex than for monopolar condensates.
As we can see, there is not only one single pair of imaginary eigenvalues 
of the stable solution approaching zero and merging with one pair of real 
unstable eigenvalues (denoted 1 in Fig.~\ref{fig:EV1gn_dipolar}(b)) 
from the excited state. 
For monopolar condensates this real pair of eigenvalues of the unstable 
solution remains the only one for increasing scattering lengths 
(see Sec.\ \ref{subsec:stabilitycoupled_monopolar}). 
By contrast, here, a second pair of eigenvalues (denoted 2 
in Fig.~\ref{fig:EV1gn_dipolar}(b)) indicating instability 
additionally forms as we follow the excited state from the bifurcation 
point to positive scattering lengths.
The eigenvectors that correspond to this pair of real eigenvalues also 
show an interesting behavior:
The stability analysis is performed in three dimensions although the 
solutions of the GPE are axisymmetric.
Therefore in the linearization around the fixed point, perturbations in $x$ 
and $y$ direction are calculated independently. 
The eigenvectors corresponding to the additional unstable eigenvalues are 
{\em not} symmetrical in $x$ and $y$, and thus break the axial symmetry 
of the fixed point solution.

The unstable branch of the calculation with two Gaussians already shows 
multiple pairs of unstable real eigenvalues. 
This suggests that the dynamics of dipolar condensates in this parameter 
region is very complex and can be described better by including even more
coupled Gaussians in the variational approach.
Indeed, the further increase of the number of variational parameters reveals 
new physical properties and phenomena also of the ground state of the 
biconcave dipolar condensate.

\subsection{Converged variational calculations with up to six coupled 
Gaussian functions}
\label{subsec:dresults:coupled}
\begin{figure}
\includegraphics[width=0.85\columnwidth]{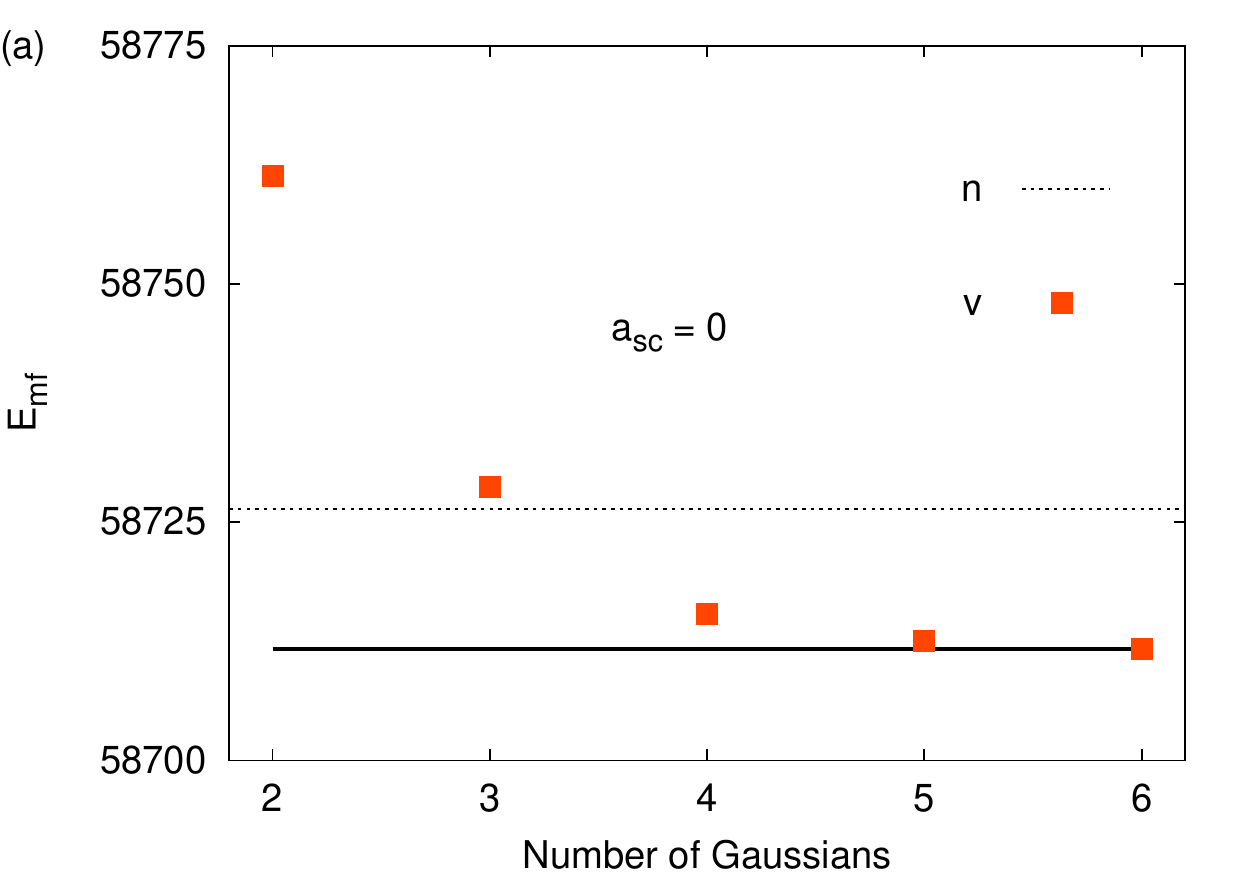}
\includegraphics[width=0.85\columnwidth]{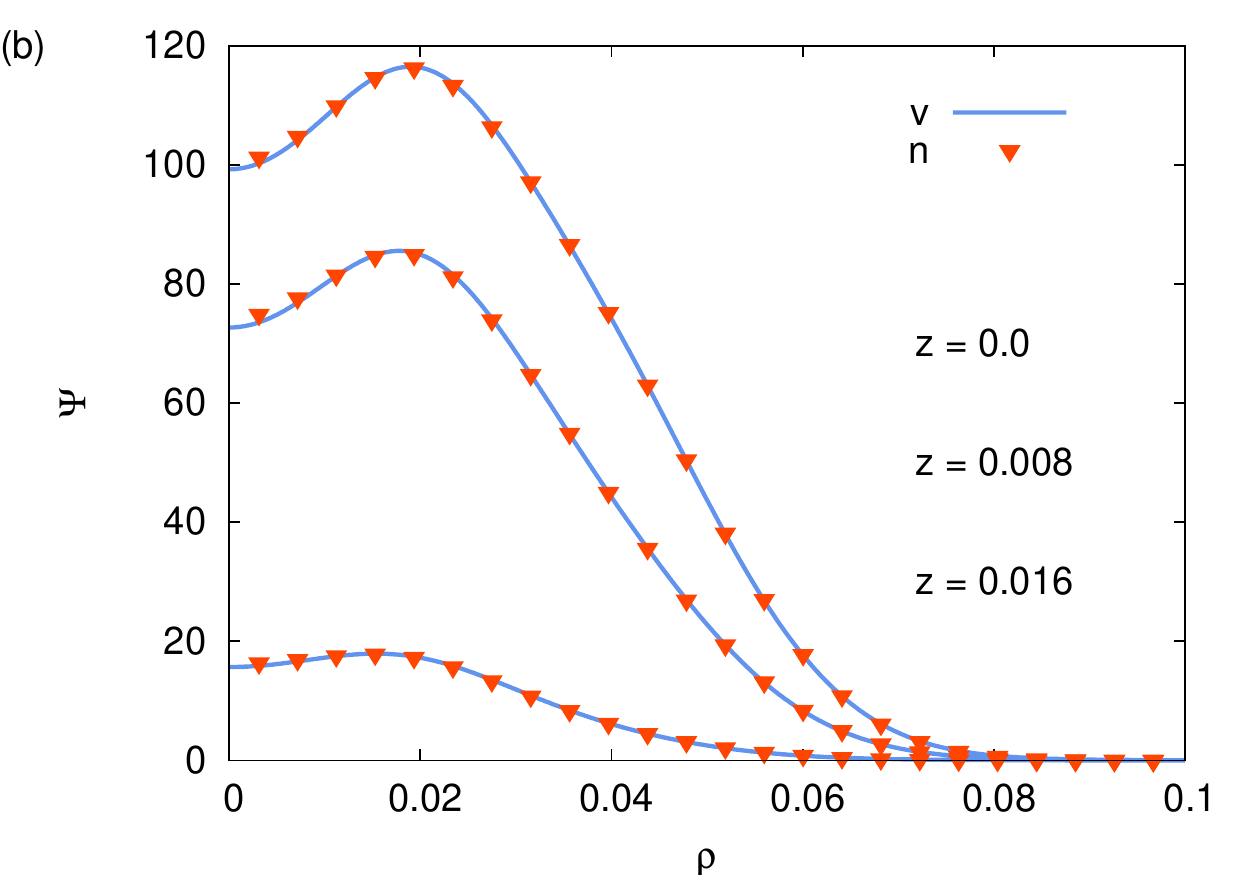}
\caption{\label{fig:compare_dipolar}
 (a) Convergence of the mean field energy (to the solid line) with increasing
 number of coupled Gaussian wave functions for $\scatt = 0$.  The mean field
 energy for four coupled functions lies already energetically lower than the
 numerical value of the lattice computation (dashed line).
 (b) The converged wave function $\psi$ as a function of the transverse
 coordinate $\varrho$ at different $z$ coordinates.
 The variational solution and the numerical lattice solution are denoted
 v and n, respectively.}
\end{figure}

Since the inclusion of two Gaussians already substantially improves the 
mean field energy, the coupling of more functions only results in a minor 
correction in the value of the mean field energy, which would not be apparent 
in, e.g., Fig.~\ref{fig:E_2g_num_dipolar}.
We therefore present in Fig.~\ref{fig:compare_dipolar}(a) the convergence 
of the mean field energy at one selected scattering length, viz.\ $\scatt=0$,
for $N=2$ to $N=6$ coupled functions. 
For other scattering lengths the convergence behavior is similar.
This example shows that as few as four coupled Gaussian functions result 
in a mean field energy which lies below the numerical solution of the
lattice calculation (dashed line). 
For five and six Gaussians, the variational solution converges to a distinct 
value (solid line). 
Note that the simplest variational solution with one Gaussian function 
is not included in the figure because the mean field energy of 
$E_{\mathrm{mf}} = 60361$ lies far outside the vertical energy scale.

In Fig.~\ref{fig:compare_dipolar}(b) for the same scattering length 
($\scatt = 0$) the converged wave function is shown at different 
$z$ coordinates. 
The wave function of the variational calculation is practically identical 
to the wave function of the numerical lattice calculation. 
Both wave functions show the characteristic biconcave shape of the condensate.

In this section, we discuss properties of the solutions obtained with five 
and six coupled Gaussians, which are qualitatively identical and 
quantitatively indistinguishable in the figures presented. 
Therefore we will omit the detailed label and refer to the converged 
variational solution simply as ``variational solution'' (v).
While the use of one and two coupled Gaussians in Sec.~\ref{subsec:dresults:2Gn}
results in two branches, one stable, one unstable, emerging in a tangent 
bifurcation, this bifurcation scenario has to be revised with the 
converged variational solution. 

\begin{figure}
\includegraphics[width=0.9\columnwidth]{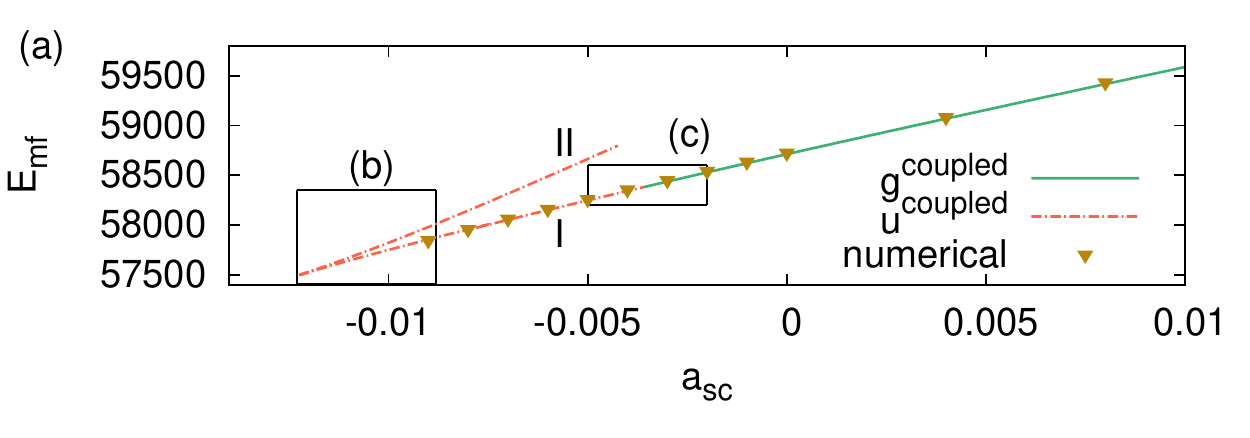}
\includegraphics[angle=-90, width=0.9\columnwidth]{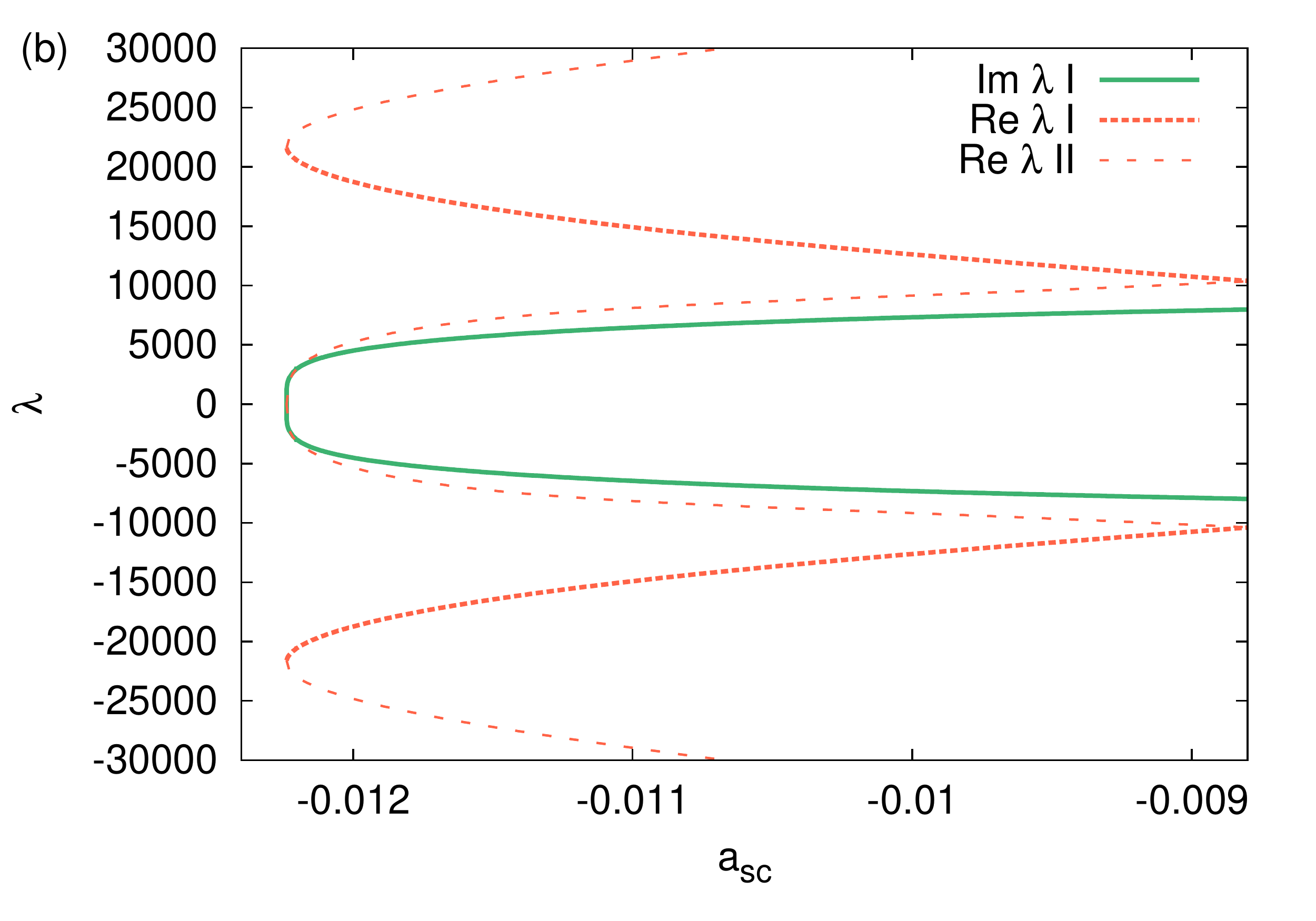}
\includegraphics[angle=-90, width=0.9\columnwidth]{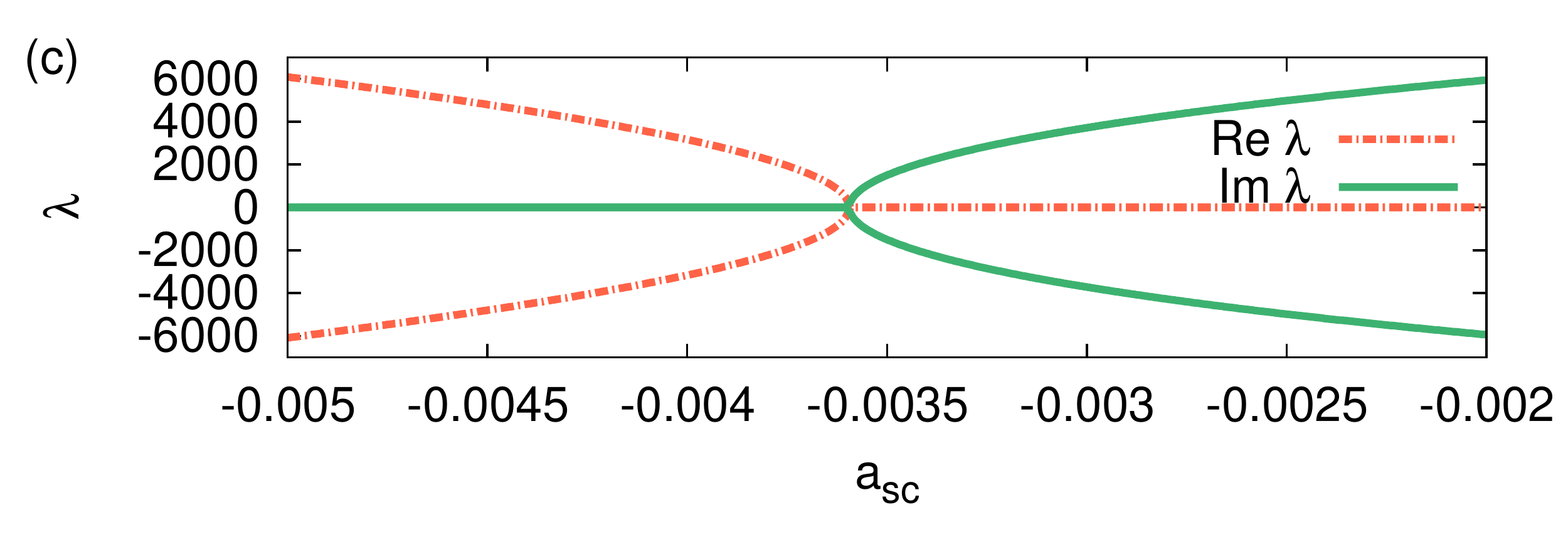}
\caption{\label{fig:E_6gn_dipolar}
 (a) Overview of the mean field energy for the bifurcation scenario
 revealed in the variational calculation. For comparison some values of
 the numerical lattice solution for the ground state are presented (see
 text for more details). The stability eigenvalues in the regions around
 the critical scattering lengths $a^{\mathrm{cr,t}}_{\mathrm{sc}} = -0.01224$
 and $a^{\mathrm{cr,p}}_{\mathrm{sc}} = -0.00359$ marked by the black frames
 in (a) are shown in (b) and (c), respectively.
 In (b) I and II denote the two merging branches.  Only the
 eigenvalues involved in the bifurcation are shown (for analysis and
 interpretation see text). Eigenvalues of the Jacobian as a function of 
 the scattering length in (c) indicate the stability change of the ground
 state in a pitchfork bifurcation.  The unstable state for
 $\scatt < \scatt^{\mathrm{cr,p}}= -0.00359$ turns into the stable ground state
 for $\scatt > \scatt^{\mathrm{cr,p}} $  in a pitchfork bifurcation.
 Subfigure (c) only shows the lowest eigenvalues, those involved in the
 stability change.}
\end{figure}

Figure~\ref{fig:E_6gn_dipolar}(a) shows an overview of the mean field energy 
for the variational solution. 
There are two important intervals of the scattering length $\scatt$, showing 
different characteristics of the variational solution with coupled Gaussian 
functions. 
These intervals of the scattering length are marked in
Fig.~\ref{fig:E_6gn_dipolar}(a).
The different line styles in Fig.~\ref{fig:E_6gn_dipolar}(a) indicate the 
stability of the solutions anticipating the results of the stability analysis. 
The numerical solution via lattice calculation and imaginary time evolution 
obtains only the ground state.
Note that the numerical simulation was carried out on a two-dimensional 
axisymmetric grid, and thus the imaginary time evolution can provide 
unstable ground states if the instability is rotational, i.e., resulting 
in an angular collapse of the condensate \cite{WilsonPRA80}.

The variational solution is able to obtain both, the stable ground state and 
stationary excited states. 
There are variational results down to a critical point 
$a_{\mathrm{cr}}^{\rm t} = -0.01224$.
To analyze the stability of the solution, we present in 
Fig.~\ref{fig:E_6gn_dipolar}(b) a stability analysis of the linearized 
dynamical equations in the interval $-0.0123 < \scatt < -0.0088$
of the scattering length [frame marked (b) in Fig.~\ref{fig:E_6gn_dipolar}(a)].

Figure~\ref{fig:E_6gn_dipolar}(b) shows at the center the typical scenario 
of eigenvalues of two branches merging in a tangent bifurcation at 
$a_{\mathrm{cr}}^{\rm t} = -0.01224$. 
A pair of purely imaginary eigenvalues of the branch denoted I 
merges at a critical point with a pair of real eigenvalues of branch II. 
Respective vanishing real or imaginary parts are not shown. 

In addition to this tangent bifurcation there is a direction in Gaussian 
parameter space, in which both branches show unstable, 
purely real eigenvalues (see Fig.~\ref{fig:E_6gn_dipolar}(b) top and bottom). 
Therefore both branches involved in the tangent bifurcation are born unstable.

The previous scenario which resulted from the calculation with one or two  
coupled Gaussians in Sec.\ \ref{subsec:dresults:2Gn} must now be revised. 
In the converged variational ansatz, the tangent bifurcation is on top of an 
unstable direction. 
The important conclusion is that there is no stable condensate in this region 
of the scattering length.

Where does the variational condensate turn stable?  To pursue this 
question we increase the scattering length to $\scatt \approx -0.00359$ 
[frame marked (c) in Fig.~\ref{fig:E_6gn_dipolar}(a)]. 
The corresponding stability analysis shows a stability change for the 
lowest eigenvalues of the ground state, which are plotted in 
Fig.~\ref{fig:E_6gn_dipolar}(c).

For scattering lengths $\scatt < \scatt^{\mathrm{cr,p}}= -0.00359$ the branch 
is unstable, indicated by the pair of real eigenvalues in 
Fig.~\ref{fig:E_6gn_dipolar}(c). 
At this bifurcation point the real eigenvalues vanish and for 
$\scatt > \scatt^{\mathrm{cr,p}}= -0.00359$ a stable ground state forms,
indicated by a pair of imaginary eigenvalues.
Figure~\ref{fig:E_6gn_dipolar}(c) shows only the respective lowest pair of 
eigenvalues which is involved in the stability change, all other eigenvalues 
are purely imaginary, and are omitted for the sake of clarity of the figure. 

The stability change of the ground state in Fig.~\ref{fig:E_6gn_dipolar}(a) 
takes place in a pitchfork bifurcation.
From left to right, one unstable branch turns stable in the bifurcation. 

In general, three branches are involved in a pitchfork bifurcation 
\cite{OttChaos}.
If the ground state of the dipolar condensate changes stability in a 
pitchfork bifurcation, two stable states should appear as stationary states 
in the variational calculation for the dipolar BEC as well.
However, for the following reasons we are probably unable to observe these 
additional stable states directly.
There are $4N$ complex Gaussian variational parameters, i.e.\ 48 real parameters for six coupled functions. 
The pitchfork bifurcation and the stability change take place in one direction 
characterized by the eigenvectors of the eigenvalues shown in 
Fig.~\ref{fig:E_6gn_dipolar}(c). 
Since an increasing number of variational parameters leads to a more and more complex parameter space 
with increasingly complex interactions between the degrees of freedom, 
it is well possible that the two stable states are limited to an extremely
tiny vicinity of the bifurcation point.

For scattering lengths greater than the bifurcation, the ground state is stable. 
At the bifurcation the system also changes from regular dynamics to chaotic dynamics, 
where for scattering lengths below the bifurcation point 
both additional stable branches may undergo several bifurcations themselves, 
immediately turning them unstable. 
Therefore it is not possible to obtain the stable branches directly via 
search for fixed points of the dynamical equations.
\begin{figure}
\includegraphics[width=0.8\columnwidth]{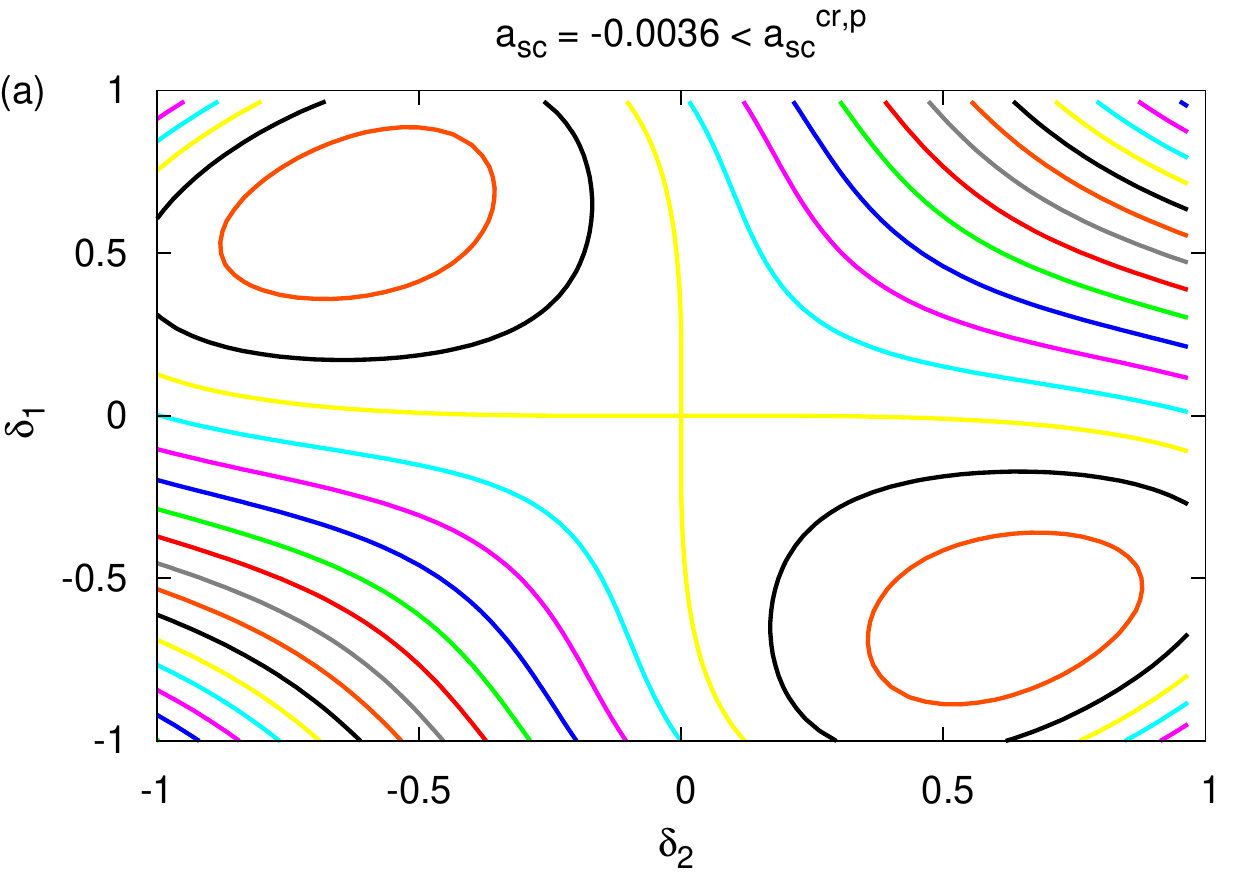}
\includegraphics[width=0.8\columnwidth]{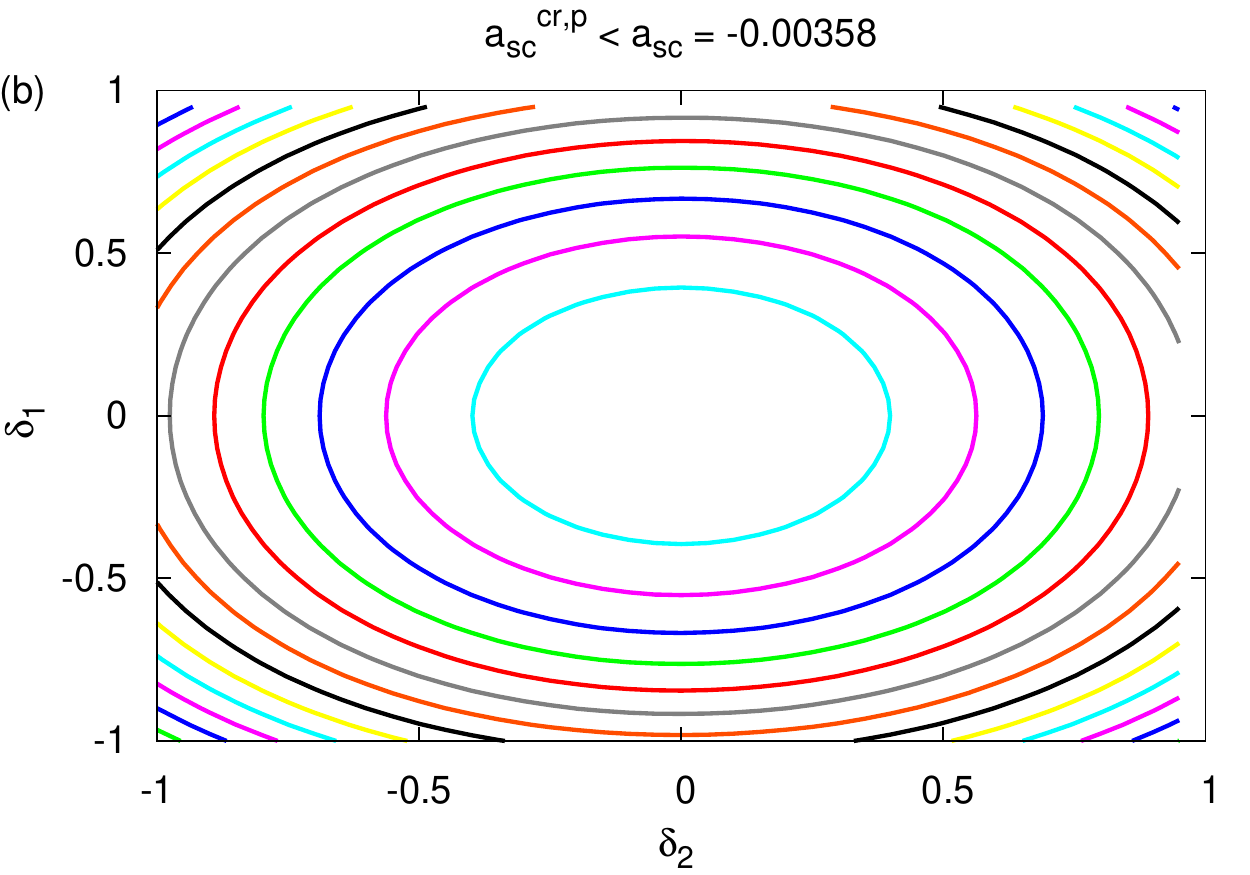}
\caption{\label{fig:_ev_6g_pitchfork_map}
 Contour plot of the mean field energy (a) for $\scatt = -0.0036$ closely
 below, and (b) for $\scatt = -0.00358$ closely above the pitchfork
 bifurcation.  The eigenvectors corresponding to the eigenvalues in
 Fig.~\ref{fig:E_6gn_dipolar}(c) linearize the vicinity of the fixed point
 ($\delta_1$ and $\delta_2$, arbitrary units).}
\end{figure}
However, it is possible to catch a glimpse of those branches, if we consider 
the linearized surroundings of the stability changing state very close to 
the bifurcation. 

The eigenvectors corresponding to the lowest eigenvalues (that show the 
stability change) linearize the vicinity of the stationary state 
$(\delta_1,\delta_2)$. 
Figure~\ref{fig:_ev_6g_pitchfork_map} shows the contour plot of the mean 
field energy of this linearization in arbitrary units for two scattering 
lengths (a) very close below and (b) very close above the bifurcation.
Above the bifurcation Fig.~\ref{fig:_ev_6g_pitchfork_map}(b) shows one 
elliptic stable state, the stable ground state.
Below the bifurcation, Fig.~\ref{fig:_ev_6g_pitchfork_map}(a) 
shows the unstable fixed point at the center.
Besides the unstable hyperbolic fixed point, there are two stable elliptic 
points at ($\pm0.5,\mp0.5$) in this vicinity linearized by the eigenvectors.
Nevertheless, Fig.~\ref{fig:_ev_6g_pitchfork_map} is limited to the 
two-dimensional plane spanned by two eigenvectors. 
If all directions of the eigenvectors of {\em all} eigenvalues are considered,
those two states are only stable in a very small interval 
$\scatt^{\mathrm{cr,p}}-\epsilon < \scatt < \scatt^{\mathrm{cr,p}}$ below the 
bifurcation point, which makes them numerically impossible to find.
Due to the numerically small value of $\epsilon$, however, the classification 
of the condensate as unstable for scattering lengths 
$\scatt < a^{\mathrm{p}}_{\mathrm{cr}} =  -0.0359$ remains true in physical terms.

If we further investigate the two eigenvectors that correspond to the pair of
eigenvalues in which the stability change occurs, we see, that the axial 
symmetry is no longer present. 
For the present trap symmetry and frequencies $(\gamma_x = \gamma_y = 3600$ 
and $\gamma_z = 25200)$ and the ansatz 
\begin{equation}
   \psi(\varrho,z)
 = \sum_{k=1}^N \ee^{\ii\left(a_\varrho^k\varrho^2+a_z^k z^2+\gamma^k \right)} \; ,
\end{equation}
all fixed points had been axisymmetric. 
The stability analysis, however, is done without any assumptions considering 
symmetry, allowing variations in both, $\delta a^k_x$ and $\delta a^k_y$ 
separately. 
Therefore, oscillations in directions which break the axial symmetry are 
allowed. 

The characteristic eigenvectors can be considered as deviations of the wave 
function $\delta\psi$ (see \cite{paper1}). 
If the eigenvector is no longer axially symmetric, i.e.,
$\delta a_x^k = -\delta a_y^k $ for all $k$, the perturbation leads to an 
asymmetric oscillation or collapse of the condensate. 
Indeed, we find this kind of eigenvectors for the lowest eigenvalues of the 
Jacobian for the variational solution of the ground state in 
Fig.~\ref{fig:E_6gn_dipolar}(c).
This behavior of the eigenvectors of the Jacobian is an indication of the 
so called ``angular collapse of dipolar BEC'' associated with the 
biconcave shape of the condensate \cite{WilsonPRA80}.

\begin{figure}
\includegraphics[angle=-90, width=0.85\columnwidth]{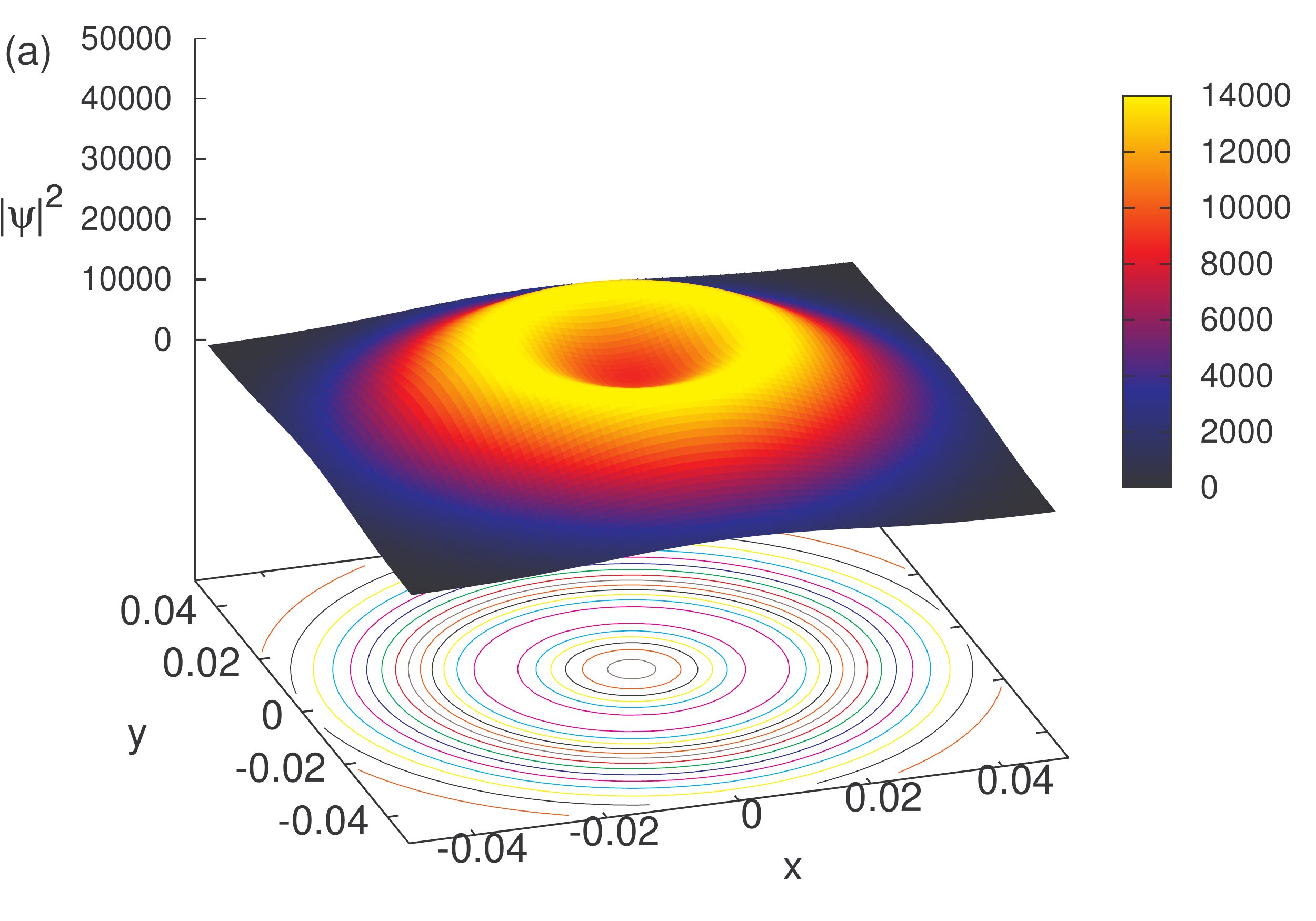}
\includegraphics[angle=-90, width=0.85\columnwidth]{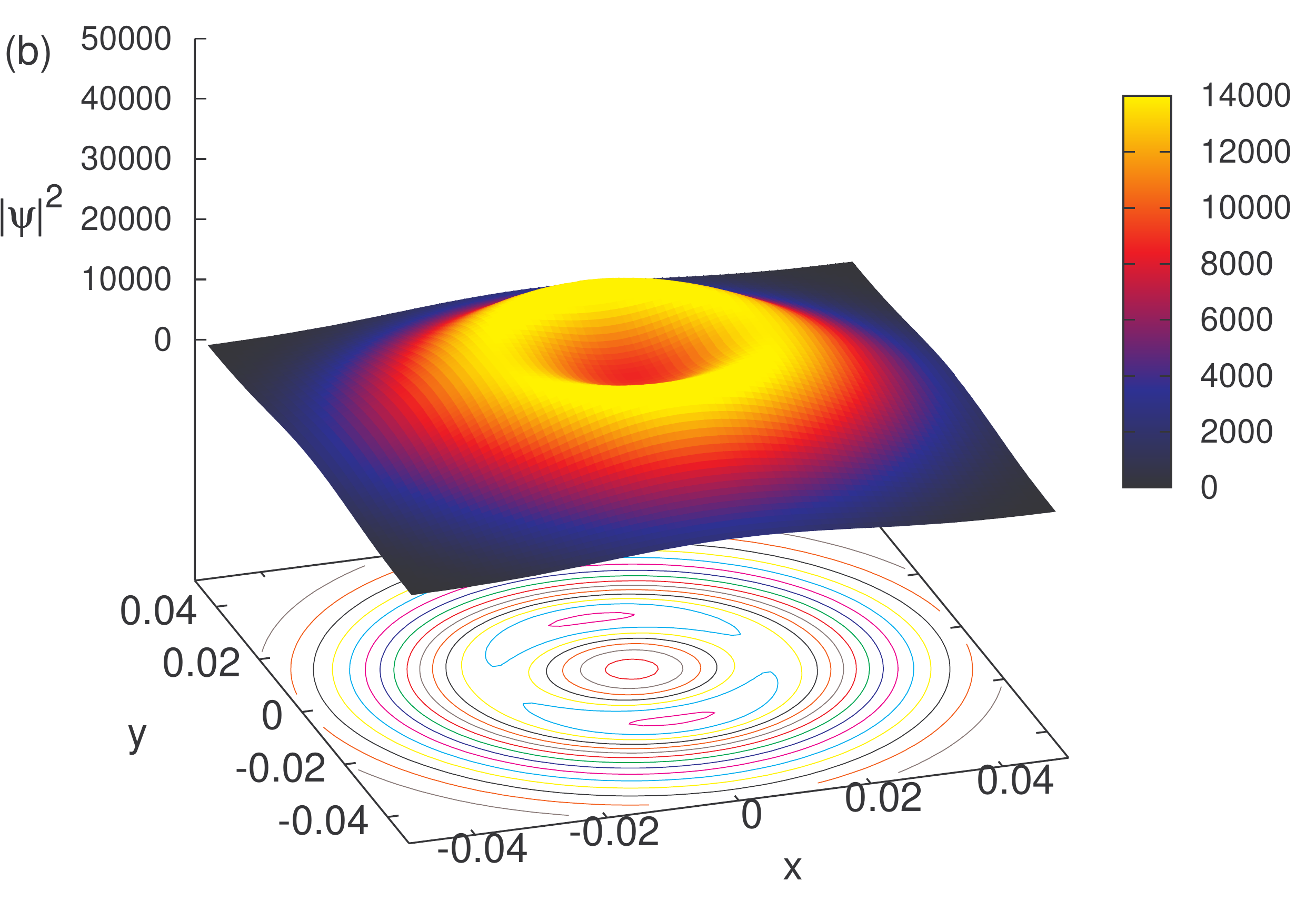}
\includegraphics[angle=-90, width=0.85\columnwidth]{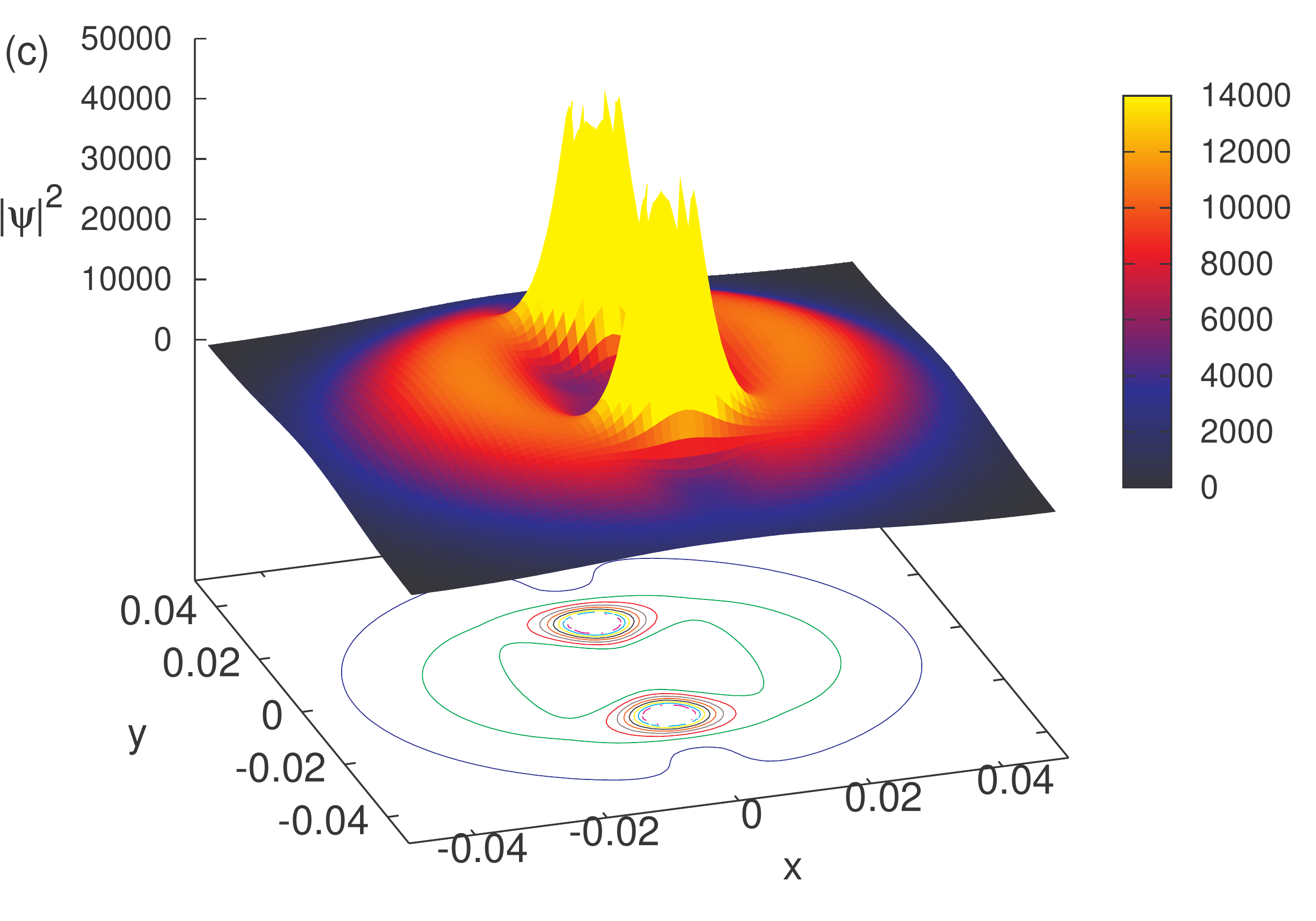}
\caption{\label{fig:realtimeevolutionunst}
 Time evolution of the perturbed particle density $|\psi|^2$ of the unstable
 stationary solution as a function of $x$ and $y$ for $z=0$ for
 (a) $t=0.0001$, (b) $t=0.005$, and (c) $t=0.006$.}
\end{figure}
This angular collapse can be observed in a time evolution of the condensate. 
We prepare the stationary wave function for a scattering length closely below 
the bifurcation, and add a deviation
\begin{align}
 \psi(\bm z) = \psi(\bm z^{\mathrm{FP}}) + \delta \psi(\delta \bm z_i)
\end{align}
in the direction of the eigenvector $i$ whose corresponding eigenvalue is 
involved in the stability change.
Figure~\ref{fig:realtimeevolutionunst} 
shows a time evolution of the particle density $|\psi|^2$ obtained by numerical integration 
of the dynamical equations. 
The time evolution of the unstable excited state clearly reveals an angular collapse of the condensate, 
the particle density concentrates on two non-axially symmetric regions as shown in Fig.~\ref{fig:realtimeevolutionunst}.
The Gaussian ansatz \eqref{defwdh:dipolartrialfunction} implies a collapse 
of the condensate with parity with respect to the $x$ and $y$ axis, 
but it may be possible in future work to modify the Gaussian functions to 
include any form of collapse without symmetry restrictions. 
A modified ansatz may even allow for an angular collapse of the condensate 
with three density peaks as reported by Wilson et al.\ \cite{WilsonPRA80}.

\section{Conclusion}
\label{sec:conclusion}
We have applied the method of coupled Gaussian wave packets to Bose-Einstein 
condensates with two different types of long-range interaction, viz.\
an attractive gravity-like $1/r$ interaction and a dipole-dipole interaction.
The mean field energy and chemical potential have been obtained as fixed
points of dynamical equations for the set of variational parameters.
As an alternative to solving the Bogoliubov equations the stability
properties of the condensates have been determined by applying methods 
of nonlinear dynamics to the linearized equations of motion.

For {\em monopolar condensates} we have shown that the additional variational 
parameters of the coupled Gaussian ansatz greatly improve the accuracy of 
the variational solution in comparison to the established single Gaussian 
ansatz.
With three coupled Gaussian functions in the trial wave function, the 
numerical mean field energy is already reproduced with an accuracy of more 
than four digits.
The solution with five Gaussians proves to be fully converged 
to the solution of the direct numerical integration of the GPE.
Furthermore, the stability properties and the bifurcation of the numerical 
solution are excellently reproduced by the coupled Gaussian ansatz.
The variational method also provides easy access to higher stability 
eigenvalues, which numerically are hard to obtain.
For monopolar condensates, the method of coupled Gaussian functions is an 
excellent and fully valid alternative to the direct numerical integration 
of the GPE.				

For {\em dipolar condensates} we have described the new phenomena revealed 
by variational solutions with an increasing number of coupled Gaussians.
The variational ansatz with multiple coupled Gaussian functions turns out to
be a full-fledged alternative to numerical lattice calculations for condensates
with dipolar interaction.
With the use of as little as five to six Gaussian functions, the variational 
solution can be considered to be fully converged.
In contrast to lattice calculations via imaginary time evolution, the 
variational ansatz also obtains excited states.
Thus the method of coupled Gaussian functions gives access to yet unexplored 
regions of the space of solutions of the GPE, and we have been able to 
clarify the theoretical nature of the collapse mechanism: 
The ground state of the condensate turns unstable in a 
pitchfork bifurcation before it finally vanishes in a tangent bifurcation.
The stability analysis indicates a further feature of the collapse mechanism: 
The condensate breaks the cylindrical symmetry on the verge of collapse, 
indicating an angular decay of the condensate.

The convergence of the variational method with Gaussians has proved to
be very fast even close to the critical scattering length, at which the
collapse of the condensate sets in. 
In future work it will be interesting to monitor the convergence of the 
ansatz in the Thomas-Fermi regime, where exact polynomial solutions of 
the wave functions are found \cite{ODellPRL84,ODel04,Ebe05}.
It will also be desirable to investigate in more detail how 
the numerical stability analysis via the Bogoliubov-de Gennes equations 
is related to the variational stability analysis.
Furthermore it will be possible to investigate real time dynamics of the decay 
of dipolar BEC and angular rotons with a modified coupled Gaussian ansatz.
The coupled Gaussian ansatz proved to be a fast and accurate alternative 
to full-numerical calculations. 
There are several interesting systems where this method can be applied 
in future work, e.g., monopolar BEC with vortices or stacks of dipolar 
condensates.

%

\end{document}